\documentclass[12pt]{iopart}

\usepackage{graphicx}  
\begin{document}

\title[Nonlinear wave interaction problems  in three dimensional case]{Nonlinear wave interaction problems  in three dimensional case}

\author{C Curr\'o $^{1}$, N Manganaro $^{1}$ and M V Pavlov$^{2,3,4}$}

\address{$^{1}$ MIFT, University of Messina, Viale Ferdinando Stagno D'Alcontres 31, 98123 Messina. Italy}
\address{$^{2}$Sector of Mathematical Physics, Lebedev Physical Institute of Russian Academy of Sciences,Leninskij Prospekt 53, 119991 Moscow, Russia}
\address{$^{3}$Department of Applied Mathematics, National Research Nuclear University MEPHI, Kashirskoe Shosse 31, 115409 Moscow, Russia}
\address{$^{4}$Department of Mechanics and Mathematics, Novosibirsk State University, 2 Pirogova street, 630090, Novosibirsk, Russia}
\eads{\mailto{ccurro@unime.it}, \mailto{nmanganaro@unime.it}, \mailto{M.V.Pavlov@lboro.ac.uk}}
\vspace{10pt}
\begin{indented}
\item[]October 2016
\end{indented}

\begin{abstract}
Three dimensional nonlinear wave interactions have been analytically described. The procedure under interest can be applied
to three dimensional quasilinear systems of first order, whose hydrodynamic reductions are homogeneous semi-Hamiltonian hydrodynamic type systems (i.e.
possess diagonal form and infinitely many conservation laws). The interaction of N waves was studied. In particular we prove that they behave like simple waves and they distort after the collision region. The amount of the distortion can be analytically computed.
\end{abstract}



\vspace{2pc}

\submitto{\NL}

\maketitle
% 
% For two-column output uncomment the next line and choose [10pt] rather than [12pt] in the \documentclass declaration
%\ioptwocol
%

\section{Introduction}
Wave interaction problems are of great interest from a theoretical point of view as well as for possible applications in applied sciences. In this  framework, a prominent role has been played by model evolution equations whose canonical structure allows for exact solutions describing relevant wave profiles as it happens for $ 2 \times 2$ hyperbolic systems involving two dependent and two independent variables. As well known, these mathematical models can be recast into a form expressing the evolution of a privileged set of field variables (Riemann invariants) along the related characteristic curves. Moreover, in the homogeneous case, the quasilinear system can be reduced to linear form through the classical hodograph transformation which, in principle, can be solved  by using the Riemann method \cite{couran,Jef}. Therefore $2 \times 2$ homogeneous models represent a prototype for determining classes of systems whose canonical structure allows for exact solutions that facilitate a full understanding of the interaction process of hyperbolic waves. Furthermore, for such a kind of models, the wave dynamics is only ruled by the behaviour of the solutions along the associated families of characteristic curves and, by making use of the special class of simple wave solutions \cite{Jef,whit}, it is possible to understand in detail the role played by different families of characteristic curves inside the interaction region.

Within such a theoretical framework, the Riemann method was extended to the nonhomogeneous case in \cite{Grund,Grund1,Grund11} and a large new classes of solutions to quasilinear systems of PDEs have been obtained in \cite{Lam1,Lam2}.
Furthermore a combined use of the hodograph method and of the differential constraints technique was considered in \cite{Jan} and quite recently was used in order to study nonlinear wave interactions  \cite{cur1,cur2} as well as discontinuous initial value problems \cite{cur3,cur4,cur5} for homogeneous and nonhomogeneous $2 \times 2$ systems.

In a different way from $2 \times 2$ models of first order PDEs, for strictly hyperbolic systems involving $N > 2$ dependent variables and two independent variables, the Riemann invariants in general do not exist so that a detailed description of wave interactions in terms of exact and closed form solutions to initial value problems is a hard task. 
In such a context the construction of solutions describing regular interactions of simple waves, the conditions for the superposition of Riemann waves in terms of initial data as well as the identification of the regions of interactions have been studied by several authors  \cite{Grund1,Grund11,Burn3,Grund2,Maj,Burn4}.

However, within the latter context a remarkable role is played by homogeneous systems of hydrodynamic type with semi-Hamiltonian structure which can be diagonalized in terms of suitable field variables which are, in fact, Riemann invariants. Actually, via the generalized hodograph method a general solution of these systems can be obtained \cite{Tsarev1, Tsarev2}. Recently, in \cite{cur6}, the approach worked out in \cite{cur1,cur2} for $2 \times 2$  hyperbolic systems, has been enlarged to these class of diagonalizable semi-Hamiltonian homogeneous hyperbolic systems in order to perform an accurate description of the associated hyperbolic wave interaction processes.

Within such a theoretical framework, here our main aim is to extend to the three dimensional case the procedure proposed in \cite{cur6} for hydrodynamic models involving two independent variables. In particular we will consider the two commuting systems \cite{maksjmp, FerTsar}
\begin{eqnarray}
&&R_{t}^{i}=(a+R^{i})R_{x}^{i}, \label{chrom1} \\  
&&R_{y}^{i}=\frac{1}{bR^{i}}R_{x}^{i} \label{chrom2} 
\end{eqnarray}
where $i=1,..,N$ and
\begin{equation}
a=\sum_{m=1}^{N} R^{m},\; \; b=\prod_{m=1}^{N} R^{m}.\label{a-b}
\end{equation}
which admit the conservation laws
\begin{equation}
b_{t}=(ab)_{x},\; \; \; a_{y}=\left( -\frac{1}{b}\right) _{x}.
\label{hydro}
\end{equation}
Hydrodynamic type systems (\ref{chrom1}) and (\ref{chrom2}) are nothing but the chromatography system written in the Lagrangian and the Euler coordinates respectively.

In passing we notice that the three dimensional two component quasilinear system (\ref{hydro}) can be written in a form of a
single quasilinear equation of second order in two alternative forms
\begin{equation*}
W_{xx}=W_{x}W_{yt}-W_{t}W_{xy},
\end{equation*}
where
\begin{equation*}
b=W_{x},\; \; \; a=\frac{W_{t}}{W_{x}}
\end{equation*}
or
\begin{equation*}
V_{yt}=V_{x}V_{xy}-V_{y}V_{xx},
\end{equation*}
where
\begin{equation*}
a=V_{x},\; \; \; b=-\frac{1}{V_{y}}.
\end{equation*}

In this paper we describe nonlinear wave interactions for a special class of solutions of the quasilinear system (\ref{hydro}) and, in fact, we extend the study of nonlinear
hyperbolic wave interactions to three dimensional case. Furthermore we  determine  the control parameters for the simple wave deformation in the interaction region in terms of the initial/boundary conditions.  The explicit evaluation of such parameters has proved to be a useful tool for describing special (soliton-like) simple wave interactions \cite{cur1,cur2,SeymourV,CurroFusco} as well as to perform quantitative measures or predictions of interest to engineering applications \cite{cur6}.

 In particular, we will focus our attention on the two commuting hydrodynamic type systems (\ref{chrom1}) and (\ref{chrom2}), whose general solution is determined in \cite{maksjmp} by the generalized hodograph method
\begin{equation}
\fl \quad x+(a+R^{k})t+\frac{y}{bR^{k}}=\frac{\partial}{\partial R^{k}}
\sum_{m=1}^{N} A_{m}(R^{m})\left(\prod_{s\neq m}^N(R^{m}-R^{s})\right)^{-1}, \quad (k=1,..,N).\label{sol}
\end{equation}
which characterizes, through (\ref{a-b}), a special class of exact solutions to (\ref{hydro}) useful for describing nonlinear $N$ wave interactions. In (\ref{sol}) $A_{m}\left(R^{m}\right)$ are arbitrary functions.

The paper is organized as follows. In section \ref{inivalue}  the general solution (\ref{sol})  is rewritten in terms of the characteristic parameters associated to the characteristic curves of  the systems (\ref{chrom1}), (\ref{chrom2}) and the explicit expression of the functions $A_m$ there involved, in terms of the initial/boundary data, is obtained. In section \ref{Nwave} a detailed analytical description of two different $N$ wave interaction problems is presented. Then, in order to validate the analytical results there obtained, in section \ref{numerical}, corresponding  numerical integrations of (\ref{hydro}) are shown. Some final comments are given in section \ref{conclusion}.

\section{Initial/boundary value problems and wave interactions}\label{inivalue}
Here our aim is to calculate the functions $ A_m $ involved in the general solution (\ref {sol}) once initial/boundary conditions are given. Such a result will be useful in the next section in order to study nonlinear $ N $ wave interactions described by (\ref{chrom1}), (\ref{chrom2}) and in turn, through (\ref{a-b}), admitted by (\ref{hydro}).

First we consider an arbitrary initial/boundary value problem.
Let $ \Gamma $ be a  smooth curve in the Euclidean space of the independent variables $ (x,y,t) $ 
\begin{equation}
\Gamma: \quad \begin{array}{llll}
x=x_0(\tau),&  
y=y_0(\tau),& 
t=t_0(\tau),&
-\infty <\tau<+\infty\end{array},\label{inicurve}
\end{equation}
and we assume the following boundary data for $ R^{m} $
\begin{equation}
 R ^{m}\left(x_0(\tau),y_0(\tau),t_0(\tau)\right) =\mathcal{R}^{m}( \tau) .\label{ini1}
\end{equation}
Following a procedure outilined in \cite{cur6}, from (\ref{sol}) evaluated on $\Gamma$, we obtain
\begin{equation}\label{ini2}
 \eqalign{ x_0(\tau)+\left(\sum_{j=1}^{N} \mathcal{R}^{j}(\tau)+\mathcal{R}^{k}(\tau)\right)t_0(\tau)+\frac{y_0(\tau) }{\mathcal{R}^{k}\left( \tau \right) }\left(
\prod_{j=1}^{N}\mathcal{R}^{j}\left( \tau \right) \right)
^{-1}\cr
=\frac{{\rm d}A_{k}}{{\rm d} R^{k}}\left( \tau \right) \left(
\prod_{j\neq k}^{N}\left( \mathcal{R}^{j}\left( \tau \right)
-\mathcal{R}^{k}\left( \tau \right) \right) \right) ^{-1}\cr    
+\left( \prod_{j\neq k}^{N}\left( \mathcal{R}^{j}\left( \tau
\right)
-\mathcal{R}^{k}\left( \tau \right) \right) \right) ^{-1}\sum_{s\neq k}^{N}\frac{%
A_{k}\left( \tau \right) }{\left( \mathcal{R}^{s}\left( \tau \right)
-\mathcal{R}^{k}\left( \tau
\right) \right) }  \cr
-\sum_{s\neq k}^{N}\frac{A_{s}\left( \tau \right) }{\left(
\mathcal{R}^{k}\left(
\tau \right) -\mathcal{R}^{s}\left( \tau \right) \right) }\left( \prod_{j\neq s}^{N}\left( \mathcal{R}^{j}\left( \tau \right) -\mathcal{R}^{s}\left( \tau
\right)
\right) \right) ^{-1}  \quad\quad\left(k=1,...N\right).  }
\end{equation}
Therefore it is straightforward to ascertain that multiplying by $\frac{{\rm d}\mathcal{R}^{k}%
}{{\rm d}\tau }$ the corresponding equation of the set (\ref{ini2}),
by taking the sum over $k=1,...N$, a further integration  allows us to express
$A_{N}\left( \tau \right) $ in terms of $A_{1}\left( \tau \right)
,....A_{N-1}\left( \tau \right) $ as follows
\begin{equation}
\fl \eqalign{ A_{N}( \tau ) =\prod_{j=1}^{N-1}\left( \mathcal{R}^{j}( \tau
) -\mathcal{R}^{N}( \tau) \right) H_{1}( \tau )+\sum_{s=1}^{N-1}\left( \prod_{j\neq
s}^{N-1}\frac{\mathcal{R}^{j}( \tau ) -\mathcal{R}^{N}( \tau )
}{\mathcal{R}^{j}( \tau ) -\mathcal{R}^{s}( \tau ) }\right)
A_{s}( \tau ) } \label{IINt}
\end{equation}
along with
\begin{equation}
\fl \eqalign{\frac{{\rm d}H_{1}}{{\rm d}\tau }=x_0(\tau)\frac{{\rm d}}{{\rm d}\tau }\left(\sum_{i=1}^N \mathcal{R}^i(\tau) \right)\nonumber + t_0(\tau)\frac{{\rm d}}{{\rm d}\tau }\left(\sum_{i\leq j}^N \mathcal{R}^i(\tau) \mathcal{R}^j(\tau)\right)-y_0(\tau) \frac{{\rm d}}{{\rm d}\tau }\left(
\prod_{i=1}^{N}\mathcal{R}^{i}\left( \tau \right) \right) ^{-1}.}
\label{IINT1}
\end{equation}

By inserting the expressions (\ref{IINt}) into relations (\ref{ini2})  for $k=1,....,N-1$,  multiplying
each equation by $\frac{{\rm d}R^{k}}{{\rm d}\tau }$,  taking the sum over
$k$ and integrating the resulting relation we are able to obtain
$A_{N-1}\left( \tau \right) $. Whereupon a further iterative
procedure gives rise to
\begin{equation}
 \eqalign{A_{N-h}\left( \tau \right) =\sum_{j=1}^{N-h-1}\left(
\prod_{m=1}^{N-h-j}\left(\mathcal{R}^{m}\left( \tau \right) -\mathcal{R}^{N-h}\left(
\tau \right) \right) \right) H_{j+h}\left( \tau \right)
+H_{N}( \tau )\cr(h=0,\dots, N-1)}\label{A-h}
\end{equation}
where $ H_{h}(\tau)\, (h=2,\dots,n) $  are obtained from
\begin{equation}
 \eqalign{\frac{{\rm d}H_{h}}{{\rm d}\tau }=\sum_{k=1}^{N-h+1}\left\{\Xi^{k}( \tau )\prod_{m=N+2-h}^{N}\left( \mathcal{R}^{m}( \tau )
-\mathcal{R}^{k}( \tau
) \right)  -H_{h-1}( \tau ) \right. \cr
\left. -\sum_{j=0}^{h-2}\left(
\prod_{m=N+2-h}^{N-j}\left( \mathcal{R}^{m}\left( \tau \right) -\mathcal{R}^{k}\left(
\tau
\right) \right) \right) H_{j}\left( \tau \right) \right\} \frac{{\rm d}\mathcal{R}^{k}}{%
{\rm d}\tau },\quad H_{0}( \tau ) =0  }\label{H2} 
\end{equation}
with
\begin{equation*}
 \Xi^k(\tau)=x_0(\tau)+\left(\sum_{j=1}^{N} \mathcal{R}^{j}(\tau)+\mathcal{R}^{k}(\tau)\right)t_0(\tau)+\frac{y_0(\tau) }{\mathcal{R}^{k}\left( \tau \right) }\left(
\prod_{m=1}^{N}\mathcal{R}^{m}\left( \tau \right) \right)
^{-1}.
\end{equation*}
Finally, from (\ref{A-h}), after some algebra, the function $ A_k $  are determined in terms of the boundary data (\ref{inicurve}), (\ref{ini1})

\begin{equation}\label{wavelets11}
 \eqalign{A_{k}( \tau) =\left(\mathcal{R}^k(\tau)\right)^3 x_0(\tau)+\left(\mathcal{R}^k(\tau)\right)^4 t_0(\tau)-\frac{y_0(\tau)}{\mathcal{R}^{k}(
\tau) } \cr
+\sum_{s=0}^{N-1}\left( -1\right) ^{s}\left( \mathcal{R}^{k}( \tau) \right) ^{s}\left\lbrace\int_{0}^{\tau} 
\sum_{j_{1}<\dots<j_{s+1}}^{N}\frac{y_0^{\prime}(\xi)}{\mathcal{R}^{j_{1}}( \xi )
\dots\mathcal{R}^{j_{s+1}}( \xi ) }\, {\rm d}\xi \right. \cr
\left.-\int_{0}^\tau\sum_{j_{1}<\dots<j_{N-s}}^{N}\mathcal{R}^{j_{1}}( \xi )
\dots\mathcal{R}^{j_{N-s}}( \xi ) \left(x_0^{\prime}(\xi)+t_0^{\prime}(\xi)\sum_{i=j_1}^N \mathcal{R}^i(\xi)\right) {\rm d}\xi\right\rbrace\, ,}
\end{equation}

\bigskip
\begin{equation} \label{wavelets22}
\eqalign{
\frac{{\rm d}A_{k}}{{\rm d}R ^{k}}( \tau) =3\left(\mathcal{R}^k(\tau)\right)^2 x_0(\tau)+4\left(\mathcal{R}^k(\tau)\right)^3 t_0(\tau)+\frac{y_0(\tau)}{\left(\mathcal{R}^{k}(
\tau)\right)^2 }  \cr
+\sum_{s=0}^{N-2}\left( -1\right) ^{s+1}\left( s+1\right) \left(
\mathcal{R}^{k}( \tau) \right) ^{s}\left\lbrace\int_{0}^{\tau} \sum_{j_{1}<\dots <j_{s+2}}^{N}\frac{y^{\prime}_0(\xi)}{\mathcal{R}^{j_{1}}\left(
\xi \right)
\dots\mathcal{R}^{j_{s+2}}\left( \xi \right) }\, {\rm d}\xi  \right. \cr
\left.-\int_{0}^\tau\sum_{j_{1}<\dots<j_{N-s-1}}^{N}\mathcal{R}^{j_{1}}( \xi )
\dots\mathcal{R}^{j_{N-s-1}}( \xi ) \left(x_0^{\prime}(\xi)+t_0^{\prime}(\xi)\sum_{i=j_1}^N \mathcal{R}^i(\xi)\right) {\rm d}\xi\right\rbrace.}
\end{equation}
In (\ref{wavelets11}), (\ref{wavelets22}) and in what follows the prime denotes the derivative with respect to the indicated argument.

Next, we make use of the general expressions (\ref{wavelets11}), (\ref{wavelets22}) in order to describe nonlinear wave interactions prescribed by the quasilinear system (\ref{hydro}).  

First of all, being solution (\ref{sol}) implicitily expressed in terms of the functions $ R^k $ which remain constant along the appropriate characteristic curves $ C^{( k) } $ of the system (\ref{chrom1}), (\ref{chrom2}),  we introduce the
characteristic parameters  $\alpha _{k}\left( x,y,t\right) $ which are  ruled by the following
equations
\begin{equation}
C^{( k) }: \cases{\frac{\partial \alpha _{k}}{\partial t }-
\left(a+R^{k}\right)\frac{\partial \alpha _{k}}{\partial x
}=0 \\
\frac{\partial \alpha _{k}}{\partial y }-
\frac{1}{b R^{k}}\frac{\partial \alpha _{k}}{\partial x
}=0, \qquad \qquad \left(k=1,...,N \right)} \label{charact}
\end{equation}
so that $\alpha _{k}\left( x,y,t\right) =$ const on its own characteristic.  Therefore, bearing in mind the properties of $ 2\times 2 $ quasilinear hyperbolic systems \cite{couran}, let us assume
\begin{equation}
R ^{k}=R ^{k}\left( \alpha _{k}\right) \quad \quad
\left(k=1,...,N\right),  \label{omegak}
\end{equation}
which allows to obtain classes of solutions representing superposition of simple waves \cite{Jef,Burn1,Pera}. Such solutions are of particular interest from a physical point of view because they cover a wide range of wave phenomena arising in gas dynamics, electrodynamics, chemical engineering, etc.  \cite{couran,whit,cur6,Kuch,Burn2}.  
Therefore, owing to (\ref{omegak}), the general solution (\ref{sol}) can be expressed in terms of the
characteristic parameters $\alpha _{k}$ as follows
\begin{eqnarray}
 x+\left(a\left(\alpha
_{1},..,\alpha _{N}\right)+R^{k}(\alpha_{k})\right)t+\frac{y }{b\left( \alpha
_{1},..,\alpha _{N}\right) R^{k}(\alpha_{k})} = \nonumber \\
=\left( \prod_{j\neq k}^{N}\left( R ^{j}\left( \alpha
_{j}\right) -R ^{k}\left( \alpha _{k}\right) \right) \right)
^{-1}\left(\frac{{\rm d}A_{k}}{{\rm d}R ^{k}}\left( \alpha _{k}\right)+\sum_{s\neq k}^{N}\frac{A_{k}\left( \alpha _{k}\right)
}{ R ^{s}\left( \alpha _{s}\right) -R ^{k}\left(
\alpha _{k}\right)
 } \right)  \nonumber \\
-\sum_{s\neq k}^{N}\frac{A_{s}\left( \alpha _{s}\right)
}{ R ^{k}\left( \alpha _{k}\right) -R ^{s}\left(
\alpha _{s}\right)  }\left( \prod_{ j\neq
s}^{N}\left( R ^{j}\left( \alpha _{j}\right) -R
^{s}\left( \alpha _{s}\right) \right) \right)
^{-1},    \label{wavelets}
\end{eqnarray}
where $k=1,..,N$ and, given a function $g\left(
R ^{1},...,R ^{N}\right)$, we denote
\begin{equation*}
g\left( \alpha _{1},..,\alpha _{N}\right) =g\left( R
^{1}\left( \alpha _{1}\right) ,...,R ^{N}\left( \alpha
_{N}\right) \right).
\end{equation*}
Now, in view of investigating initial/boundary value problems associated to (\ref {chrom1}) and (\ref{chrom2})
(i. e. to (\ref {charact})), although the qualitative features apply to the general case, for simplicity we normalize $\alpha _{k}$ as follows
\begin{equation}
x =t=0,\quad \quad \quad \alpha _{k}\left( 0,y ,0\right) =y
\quad \quad \left(k=1,...,N \right)  \label{normal}
\end{equation}
so that, taking (\ref{omegak}) into account, the initial/boundary data (\ref{inicurve}), (\ref{ini1}) for
$R ^{k}$ take the form
\begin{equation}
R ^{k}\left( 0, y ,0\right) =\mathcal{R}^{k}\left( y \right),
\label{inidata}
\end{equation}
and, from (\ref{wavelets11}), (\ref{wavelets22}), the arbitrary functions $ A_k (\alpha_k)$ as well as $\frac{{\rm d}A_{k}}{{\rm d}R ^{k}}
( \alpha _{k}) $ specialize as follows

\begin{equation}\label{wavelets1}
\fl\eqalign{A_{k}\left( \alpha _{k}\right) =-\frac{\alpha _{k}}{\mathcal{R}^{k}\left(
\alpha_{k}\right) }+ \cr
+\sum_{s=0}^{N-1}\left( -1\right) ^{s}\left( \mathcal{R}^{k}\left( \alpha
_{k}\right) \right) ^{s}\int_{0}^{\alpha _{k}}\left( 
\sum_{j_{1}<\dots<j_{s+1}}^{N}\frac{1}{\mathcal{R}^{j_{1}}\left( \xi \right)
...\mathcal{R}^{j_{s+1}}\left( \xi \right) }\right) {\rm d}\xi  }
\end{equation}
\bigskip
\begin{equation} \label{wavelets2}
\fl \eqalign{
\frac{{\rm d}A_{k}}{{\rm d}R ^{k}}\left( \alpha _{k}\right) =\frac{\alpha _{k}}{%
\left( \mathcal{R}^{k}\left( \alpha _{k}\right) \right) ^{2}}+  \cr
+\sum_{s=0}^{N-2}\left( -1\right) ^{s+1}\left( s+1\right) \left(
\mathcal{R}^{k}\left( \alpha _{k}\right) \right) ^{s}\int_{0}^{\alpha
_{k}}\left( \sum_{j_{1}<\dots<j_{s+2}}^{N}\frac{1}{\mathcal{R}^{j_{1}}\left(
\xi \right)
...\mathcal{R}^{j_{s+2}}\left( \xi \right) }\right) {\rm d}\xi.}
\end{equation}

Therefore, once the initial/boundary data  (\ref{inidata}) are specified,
insertion of (\ref{wavelets1}) and (\ref{wavelets2})  into (\ref{wavelets}) allows us to
investigate in detail the evolution of the resulting wave pulses. Finally we remark that, although smooth initial
data are prescribed, the solution of quasilinear hyperbolic systems  are subject to nonlinear breakdown  \cite{Jef,whit} so that our analysis will be valid until the blow up of the solution along the characteristic curves $ C^{(k)} $ does not occur.  More precisely, if the initial/boundary data are choosen small enough on a region of the $ (y,t)- $plane , then there exists a space interval $ [0,L] $ in which the gradient catastrophe for solutions of the hydrodynamic type system (\ref{chrom1})-(\ref{chrom2})  does not occur \cite{Jan,Grund2}.

Under this assumption it is possible to select the initial/boundary data (\ref{inidata}) in such a way that every characteristic $ C^{(h)} $ has a tangent plane with inclination ( measured with respect to the positive direction of the $ y- $axis ) smaller than any characteristic of the family $ C^{(k)}\,(h<k) $ \cite{couran,Jan}. In such a case, owing to the invariance of $ R^k $  along the associated characteristic curve $ C^{(k)} $, it is straightforward to ascertain that if we consider initial/boundary data (\ref{inidata}) with compact support then the $ (x,y,t) -$space of indipendent variables  will be divided into disjoint regions of constant states, simple waves and collision regions \cite{Grund2,Burn4}. Within this analytical framework, in the next section, two different interaction processes will be described and the resulting  regions highlighted.

\section{$N$ wave interactions }\label{Nwave}
The aim of this section is to give an exact quantitative description of nonlinear $N$ wave interactions prescribed by (\ref{wavelets}) along with  (\ref{wavelets1}) and (\ref{wavelets2}). 

\begin{figure}[tbp]
$
\begin{array}{c@{\hspace{1in}}c}
\centerline{\includegraphics[width=0.6 \textwidth]{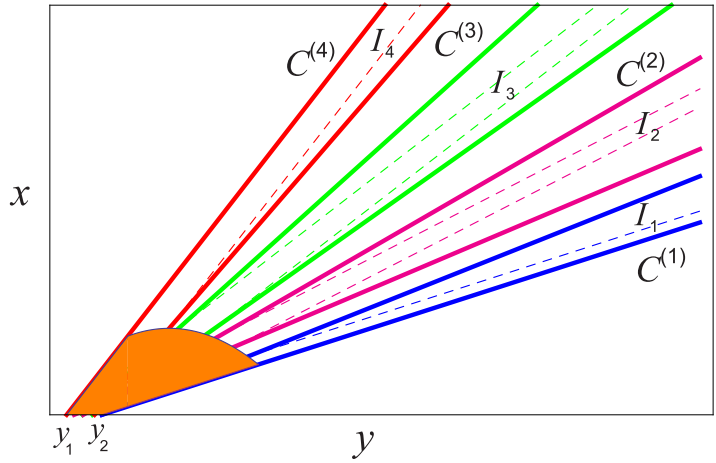}}
\end{array}
$
\caption{Qualitative behaviour in the $\left(x,y \right) -$plane of the separation process generated by initial/boundary data  (\ref{rho_k}) for \ $%
\mathcal{R} ^{j}\left(y\right) $. The
colored (orange) region represents the interaction region. $I_{k} $ ($ k=1,..4$) are the simple wave regions. The dashed lines would
correspond to the choice $\phi _{j}\left(y
\right) =\overline{\phi }_{j}$ for $j\neq k$ and point out
that the emerging pulses are distorted by the interaction process.}
 \label{Dinteract1}
\end{figure}

\begin{figure}[tbp]
$
\begin{array}{c@{\hspace{1in}}c}
\centerline{\includegraphics[width=0.7\textwidth]{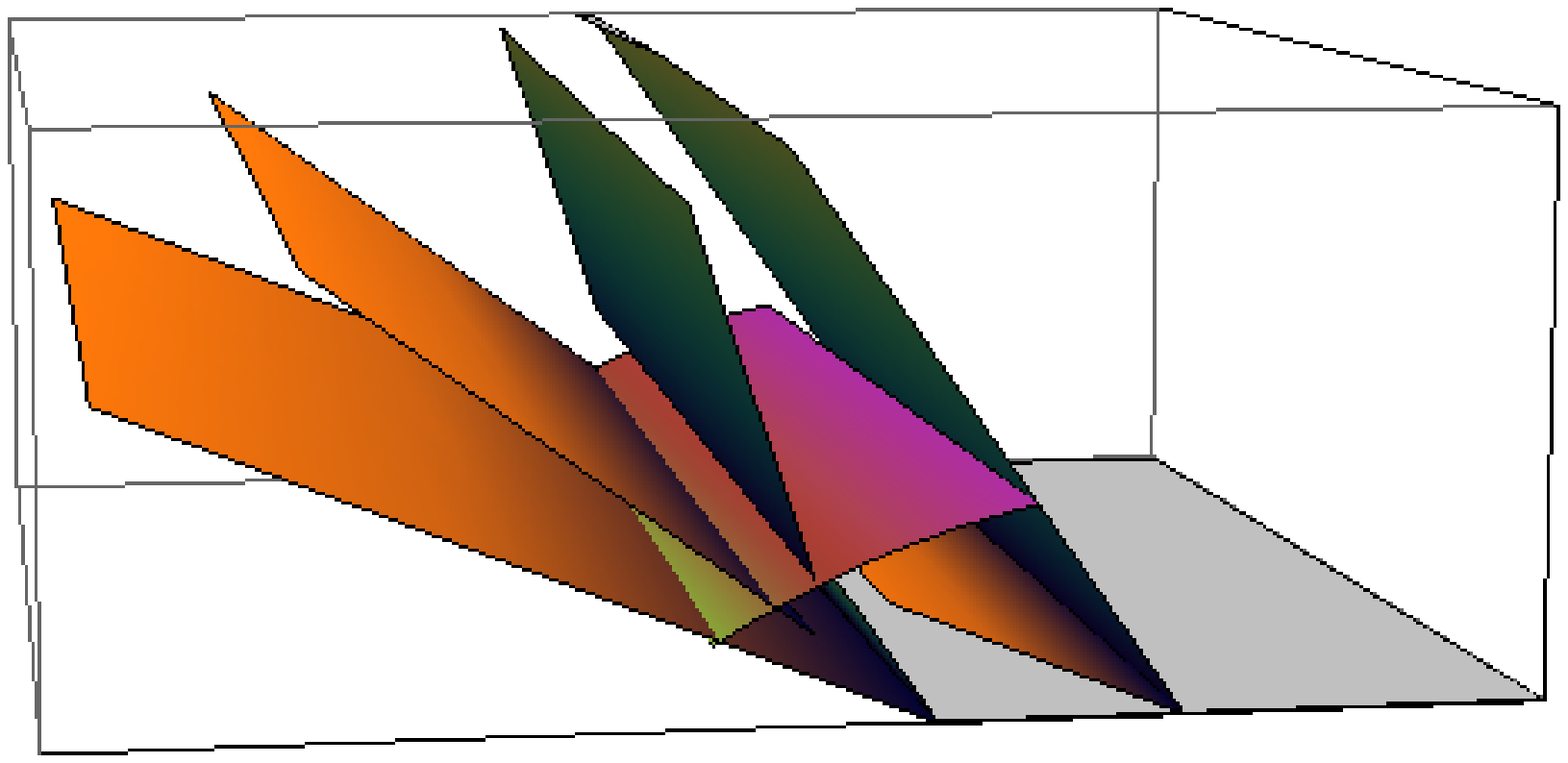}}
\put(-60,90){\fontsize{12}{12}$x $}
\put(-150,2){\fontsize{12}{12} $
y_{1}$} \put(-200,-1){\fontsize{12}{12} $y_{2}$}\put(-270,140){%
\fontsize{12}{12}$I_{1}$} \put(-355,115){\fontsize{12}{12}$I_{2}$}
\put(-335,-5){\fontsize{12}{12}$y$}
\put(-110,40){\fontsize{12}{12}$t$}
\put(-65,10){\fontsize{12}{12}$0$} &
\end{array}$
\caption{Qualitative behaviour in the $\left( x
,y,t
\right) -$space of the interaction process generated by initial/boundary data (\ref{DATA_3}) for  $%
\mathcal{R} ^{j}\left( y \right) $. $I_{1}, I_{2}$
 are the simple wave regions. The emerging pulses are distorted by the interaction process.
} \label{Dinteract3}
\end{figure}

Therefore let us  consider the initial/boundary data
\begin{equation} \label{rho_k}
\eqalign{
R^{k}(0,y,0 ) =\mathcal{R}^{k}( y )
=\left\{
\begin{array}{l}
\phi_{k}( y ) \quad\quad y_{1}\leq y\leq y_{2}  \cr
\cr
\overline{\phi }_{j}\quad\quad\quad \mbox{ otherwise}
\end{array}
\right.   \cr
\phi_{k}( y_{1}) =\phi_{k}( y_{2}) =\overline{\phi }
_{k}, \quad \quad \left(k=1,2,...,N\right)  }
\end{equation}
where $y_{1}<y_{2}$ are real numbers, $\phi _{j}( y) $ are
smooth functions and $\overline{\phi }_{j}$ $\neq 0$ are arbitrary
constants. In passing we notice that the initial/boundary value problem (\ref{rho_k}) describes $N$ waves localized at $ x=t=0 $ in the interval $\left[y_1,y_2\right]$  which, according to (\ref{charact}), propagate through regions which are adjacent to  constant states so that the pulses in point after a finite space $ x $ separate themselves and become in fact simple waves. It results that in the $( x ,y,t)-$space there are
$ N $ distinct simple wave regions $ I_{k} $ (see figure \ref{Dinteract1} and figure \ref{Dinteract3} where, for the sake of simplicity,  the cases $ N=4$ and $ N=2 $ are shown, respectively, in the plane $t=$constant  and in $(x,y,t)-$space) where each characteristic parameter $
\alpha _{k}$ can be explicitly calculated so that, once the
initial/boundary data (\ref{rho_k}) are specified, by using
(\ref{wavelets}) the behaviour of the emerging simple waves can be
fully investigated.

\begin{figure}[tbp]
\centerline{\includegraphics[width=0.60\textwidth]{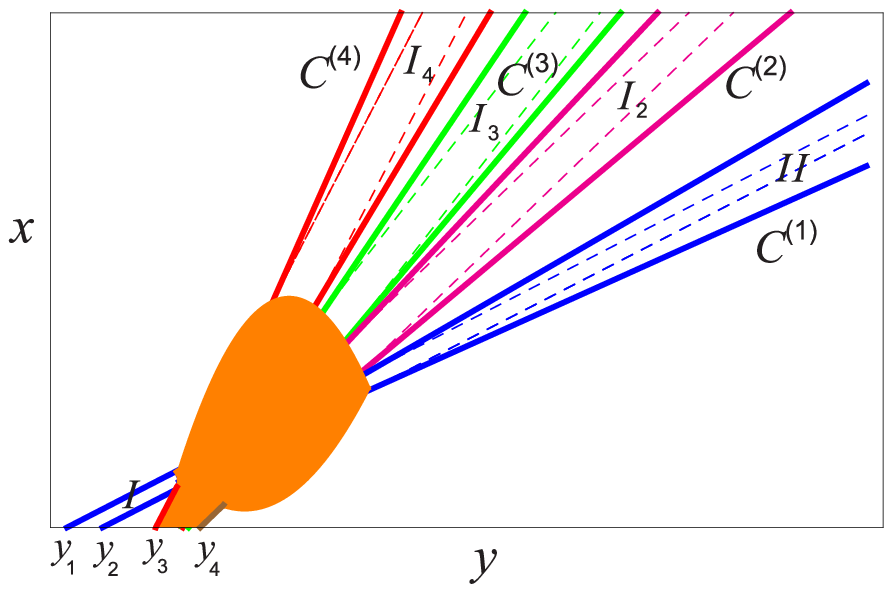}}
\caption{Qualitative behaviour in the $\left( x
,y\right) -$plane of the interaction process generated by initial/boundary data (\ref{DATA_3}) for \ $%
 \mathcal{R} ^{j}\left( y \right) $. The
colored (orange) region represents the interaction region. $I, II,
I_{2},I_{3},I_{4}$ are the simple wave regions. The dashed lines would
correspond to the choice $\psi _{j}\left( y
\right) =\overline{\psi }_{j}$ for $j\neq k$ and point out
that the emerging pulses are distorted by the interaction process.
} \label{Dinteract2}
\end{figure}

In particular, for a fixed value $k$, the simple wave travelling
along the characteristic curve  $C^{( k) }$ corresponds to the region
\begin{equation}
\eqalign{
 I_{k}:\, R^{k}=\phi _{k}\left( \alpha _{k}\right), \quad \quad R^{j}=\overline{
\phi }_{j}, \cr 
\mbox{with} \quad y_{1}\leq \alpha _{k}\leq y_{2},\quad
 \alpha _{j}\leq y_{1}\; (j<k), \quad  \alpha _{j}\geq y_{2}\; ( j>k)
\label{SIMPLE_k}}
\end{equation}
and from (\ref{wavelets}), (\ref{wavelets1}), (\ref{wavelets2}) and (\ref{rho_k}), after some algebra, we obtain the characteristic
wave parameters in each simple wave region $ I_{k} $
\begin{equation}
y+b\left(\alpha_{k}\right)\phi_{k}\left(\alpha_{k}\right)\lbrace x+\left[ a\left(\alpha_{k}\right)+\phi_{k}\left(\alpha_{k}\right)\right]t \rbrace  =\alpha
_{k}+\Lambda _{k}\left( \alpha _{k}\right)
\label{I}
\end{equation}
where $\Lambda _{k}\left( \alpha _{k}\right) $
are given by
\begin{equation}
\eqalign{
\Lambda _{k}\left( \alpha _{k}\right) &=\left(\phi_{k}\left(\alpha_{k}\right)\right)^{2}
\left\lbrace\sum_{j>k}^N\frac{\Omega_{j}(\alpha_{k})-\Omega_{j}(y_{2})}{\left(\overline{\phi}_{j}-\phi_{k}(\alpha_{k})\right)^{2}}\left(\prod_{j\neq l\neq k}^N\frac{\overline{\phi}_{l}}{\overline{\phi}_{j}-\overline{\phi}_{l}}\right)+\right.\medskip\cr
&\left.+\sum_{j<k}^N\frac{\Omega_{j}(\alpha_{k})}{\left(\overline{\phi}_{j}-\phi_{k}(\alpha_{k})\right)^{2}}\left(\prod_{j\neq l\neq k}^N\frac{\overline{\phi}_{l}}{\overline{\phi}_{j}-\overline{\phi}_{l}}\right)\right\rbrace,\quad\quad\quad k=1,...,N}
\label{interpro}
\end{equation}
with
\begin{equation}
\Omega_{j}(\tau)=\int_{y_1}^{\tau}\prod_{l}^N\left(1-\frac{\overline{\phi}_{j}}{\phi_{l}(\xi)}\right) {\rm d}\xi.
\end{equation}
The interaction terms $\Lambda _{k}\left( \alpha _{k}\right) $
represent a quantitative ''measure'' of the distortion of the
simple wave travelling along $C^{(k) }$ which depends on the initial/boundary data (\ref{rho_k}) and vanish  if there is only the localized pulse travelling along $C^{(k)}$, that is
\begin{equation}
\Lambda_{k}(\alpha_{k})=0\quad \mbox{if}\quad \phi_{j}(y)=\overline{\phi}_{j}\quad \forall y,\quad j\neq k.
\end{equation}
Next we are interested in studying the interaction between $N-1$ waves and a single pulse initially localized in disjoint intervals. Therefore we consider the following initial/boundary data
\begin{equation}\label{DATA_3}
 \fl \eqalign{
R^{k}( 0,y,0) =\mathcal{R}^{k}(y)
=\left\{
\begin{array}{l}
\psi _{k}(y) \quad y_{3}\leq y\leq y_{4}  \cr
\cr
\overline{\psi }_{k}\quad \mbox{ otherwise}
\end{array}
\right.  \psi _{k}( y_{3}) =\psi _{k}( y_{4}) =\overline{\psi }%
_{k}\,\left( k\geq 2\right)   \cr
\cr
R^{1}(0,y,0) =\mathcal{R}^{1}( y)
=\left\{
\begin{array}{l}
\psi _{1}( y) \quad y_{1}\leq y \leq y_{2} \cr
\cr
\overline{\psi }_{1}\quad \mbox{ otherwise}
\end{array}
\right. \quad \psi _{1}( y_{1}) =\psi _{1}( y_{2}) =%
\overline{\psi }_{1} . }
\end{equation}
where $y_{1}<y_{2}<y_{3}<y_{4}$ denote real constants, $\psi
_{k}(y) $ are smooth functions and $\overline{\psi
}_{k}\neq 0$ are
arbitrary constants. In such a case the pulse travelling along the characteristic curve $%
C^{(1) }$ traverses region $ I $
where it is a
simple wave, interacts with the pulses travelling along the $%
C^{(k)}$ ($ k\geq 2 $)  characteristic curves and emerges in
region $ II $ as simple wave (see figure \ref{Dinteract2}  for $ N=4  $).

\begin{figure}[tbp]
$
\begin{array}{c@{\hspace{1in}}c}
\centerline{\includegraphics[width=0.6\textwidth]{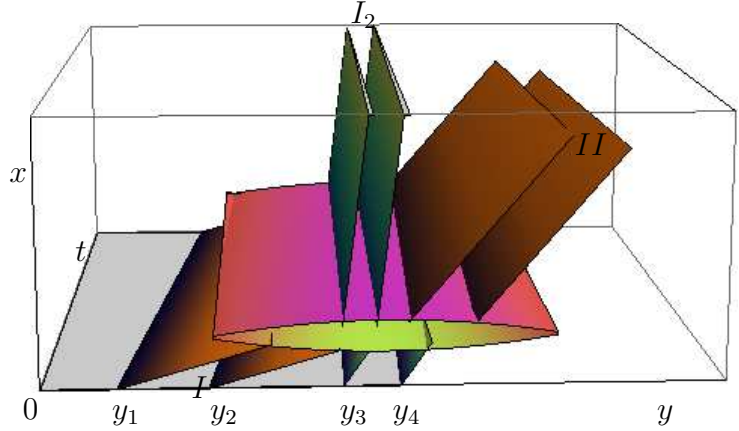}}
\put(-365,80){\fontsize{12}{12}$x $}
\put(-340,50){\fontsize{12}{12}$t $}
\put(-120,-10){\fontsize{12}{12}$y $}
\put(-155,90){\fontsize{12}{12}
$II$} \put(-300,-2){\fontsize{12}{12} $I$} \put(-330,-10){\fontsize{12}{12} $y_{1}$} \put(-293,-10){\fontsize{12}{12} $y_{2}$} \put(-240,140){
\fontsize{12}{12}$I_{2}$}
\put(-240,-10){\fontsize{12}{12}$y_{3}$}
\put(-220,-10){\fontsize{12}{12}$y_{4}$}
\put(-360,-10){\fontsize{12}{12}$0$} &
\end{array}$
\caption{Qualitative behaviour in the $\left( x
,y,t
\right) -$space of the interaction process generated by initial/boundary  data (\ref{DATA_3}) for  $%
\mathcal{R}^{j}\left( y \right) $ ($ N=2 $). The
colored (green) region represents the interaction region. $I, II,
I_{2}$ are the simple wave regions. The emerging pulses are distorted by the interaction process.
} \label{Dinteract4}
\end{figure}

In the $( x,y,t)-$space there are $ N+1 $  distinct
simple wave regions where each characteristic parameter $\alpha
_{k}$ can be explicitly calculated (see figure \ref{Dinteract4} for  $ N=2 $).

In particular from (\ref{wavelets}), (\ref{wavelets1}) and (\ref{wavelets2}),  along with
(\ref{DATA_3}), we obtain the following expressions for the characteristic
wave parameters in each simple wave region

\begin{description}
\item  REGION $ I $
\begin{equation}
\eqalign{
R^{1}=\psi _{1}\left( \alpha _{1}\right) ,\quad R^{k}=\overline{
\psi }_{k}, \cr
y+ b\left(\alpha_{1}\right)\psi_{1}\left(\alpha_{1}\right)\lbrace x +\left[a\left(\alpha_{1}\right)+ \psi _{1}\left( \alpha _{1}\right)\right]t\rbrace =\alpha_{1}  \cr
\mbox{with} \quad y_{1}\leq
\alpha _{1}\leq y_{2},\quad y_{2}\leq \alpha _{k}\leq y_{3}\;  \; \left( k\geq 2\right)}
\label{REGION_I}
\end{equation}
\item  REGION $ II $
\begin{equation}
\eqalign{
R^{1}=\psi _{1}\left( \alpha _{1}\right) ,\quad R^{k}=\overline{\psi }_{k}, \cr
y+b\left(\alpha_{1}\right)\psi_{1}\left(\alpha_{1}\right) \lbrace x +\left[a\left(\alpha_{1}\right)+ \psi _{1}\left( \alpha _{1}\right)\right]t\rbrace =\alpha_{1}+\Theta_{1}\left(\alpha_{1}\right) \cr
\mbox{with} \quad y_{1}\leq
\alpha _{1}\leq y_{2},\quad  \alpha _{k}\geq y_{4}\quad \left( k\geq 2\right)}
\label{Region_IV}
\end{equation}
\item  REGION $ I_{k} \quad (k\geq 2)$
\begin{equation}
\eqalign{
R^{k}=\psi _{k}\left(
\alpha _{k}\right) ,\quad R^{j}=\overline{\psi }_{j}\quad (j\neq k),\cr
y+ b\left(\alpha_{k}\right)\psi_{k}\left(\alpha_{k}\right)\lbrace x +\left[a\left(\alpha_{k}\right)+ \psi _{k}\left( \alpha _{k}\right)\right]t\rbrace =\alpha_{k}+\Theta_{k}\left(\alpha_{k}\right) \cr
\mbox{with} \quad y_{3}\leq \alpha_{k}\leq y_{4},\quad \alpha _{1}\leq y_{1},\quad \alpha _{j}\geq y_{4} \; \; (j>k),\quad \alpha _{j}\leq y_{3} \; \; (j<k)
\label{Region_II}}
\end{equation}
\end{description}

\noindent In (\ref{Region_IV}), (\ref{Region_II}) the functions
$\Theta _{k}\left( \alpha _{k}\right) $ measure the $C^{(
k) }$ wave parameters distortion due to the
interaction with the pulses travelling along the $C^{(
j) } \, \left( j\neq k\right)$ characteristics and
are given by
\begin{equation} \label{interpro_III}
\Theta _{1}\left( \alpha _{1}\right) =\left(\psi_{1}(\alpha_{1})\right)^{2}\sum_{j>1}^N\frac{\left(\overline{\psi}_{j}-\overline{\psi}_{1}\right)\Delta_{j}(y_{4})}{\overline{\psi}_{1}\left(\overline{\psi}_{j}-\psi_{1}(\alpha_{1})\right)^{2}}\prod_{1\neq l \neq j}^N\left(\frac{\overline{\psi}_{l}}{\overline{\psi}_{j}-\overline{\psi}_{l}}\right),
\end{equation}
\vspace{0.1cm}
\begin{equation} \label{interpro_IV}
\fl \eqalign{
\Theta _{k}\left( \alpha _{k}\right) =\left(\psi_{k}(\alpha_{k})\right)^{2}\left\lbrace \frac{\left(\overline{\psi}_{1}-\overline{\psi}_{k}\right)\Delta_{1}}{\overline{\psi}_{k}\left(\overline{\psi}_{1}-\psi_{k}(\alpha_{k})\right)^{2}}\right.+\sum_{j>k}^N\frac{\Delta_{j}(\alpha_{k})-\Delta_{j}(y_{4})}{\left(\overline{\psi}_{j}-\psi_{k}(\alpha_{k})\right)^{2}}\prod_{1\neq l \neq j,k}^N\left(\frac{\overline{\psi}_{l}}{\overline{\psi}_{j}-\overline{\psi}_{l}}\right)+ \medskip  \cr
+\left. \sum_{1<j<k}^N\frac{\Delta_{j}(\alpha_{k})}{\left(\overline{\psi}_{j}-\psi_{k}(\alpha_{k})\right)^{2}}\prod_{1\neq l \neq j,k}^N\left(\frac{\overline{\psi}_{l}}{\overline{\psi}_{j}-\overline{\psi}_{l}}\right)\right\rbrace, \quad\quad k=2,...,N}
\end{equation}
with
\begin{equation}
\fl \Delta_{1}=\int_{y_{1}}^{y_{2}}\left(1-\frac{\overline{\psi}_{1}}{\psi_{1}(\xi)}\right) {\rm d}\xi \, ,\quad  \Delta_{j}(\tau)=\int_{y_3}^{\tau}\prod_{l>1}^N\left(1-\frac{\overline{\psi}_{j}}{\psi_{l}(\xi)}\right){\rm d}\xi\; \; \left(j\geq 2\right).
\end{equation}
We notice that the pulses generated by the localized data $R^{j}( 0,y,0) \, \left(
j\geq 2\right) $, interact
with the fastest simple wave travelling along the $C^{(1)} $ characteristic curve and emerge,
after interaction, with altered profiles. The parameters distortion
$\Theta _{k}\left( \alpha _{k}\right) $  are given in terms of the initial/boundary data and determine a quantitative
measure of this alteration. Therefore the resulting wave's distortion depend strongly by the set of initial/boundary values under interest.   

\begin{figure}[tbp]
$
\begin{array}{c@{\hspace{1in}}c}
\begin{array}{cc}
\includegraphics[width=0.50\textwidth]{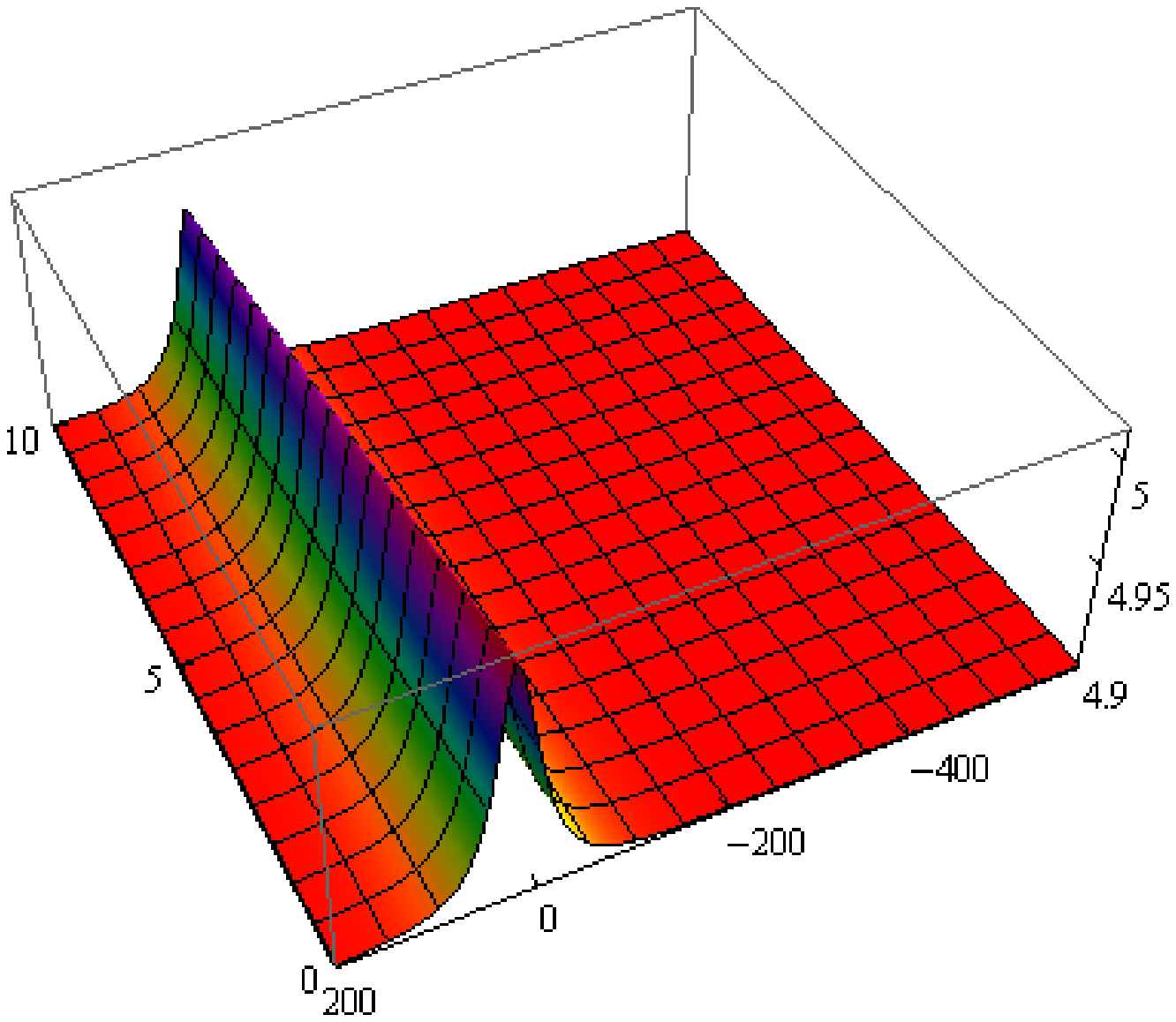} & \includegraphics[width=0.4\textwidth]{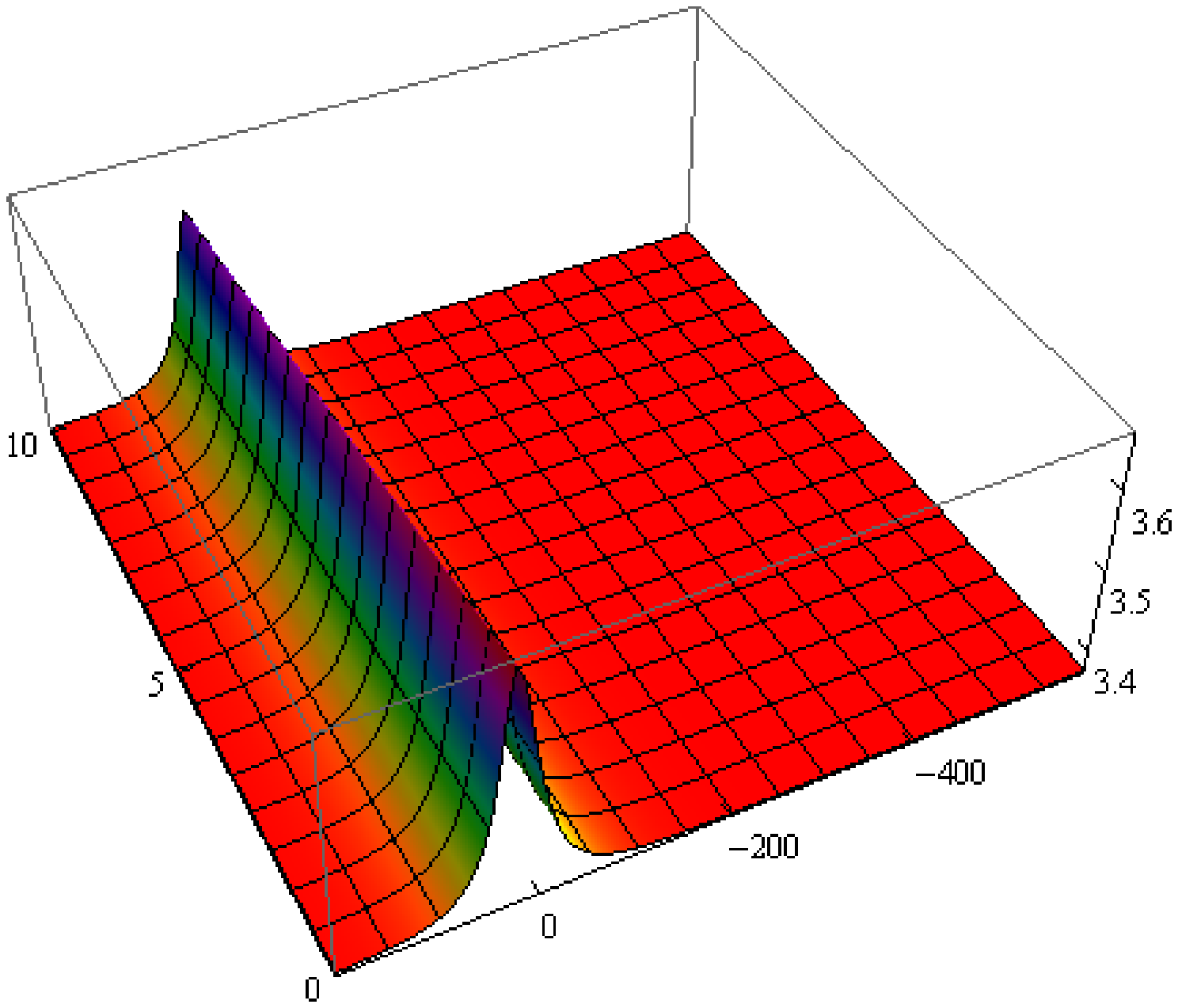} \\ 
\includegraphics[width=0.50\textwidth]{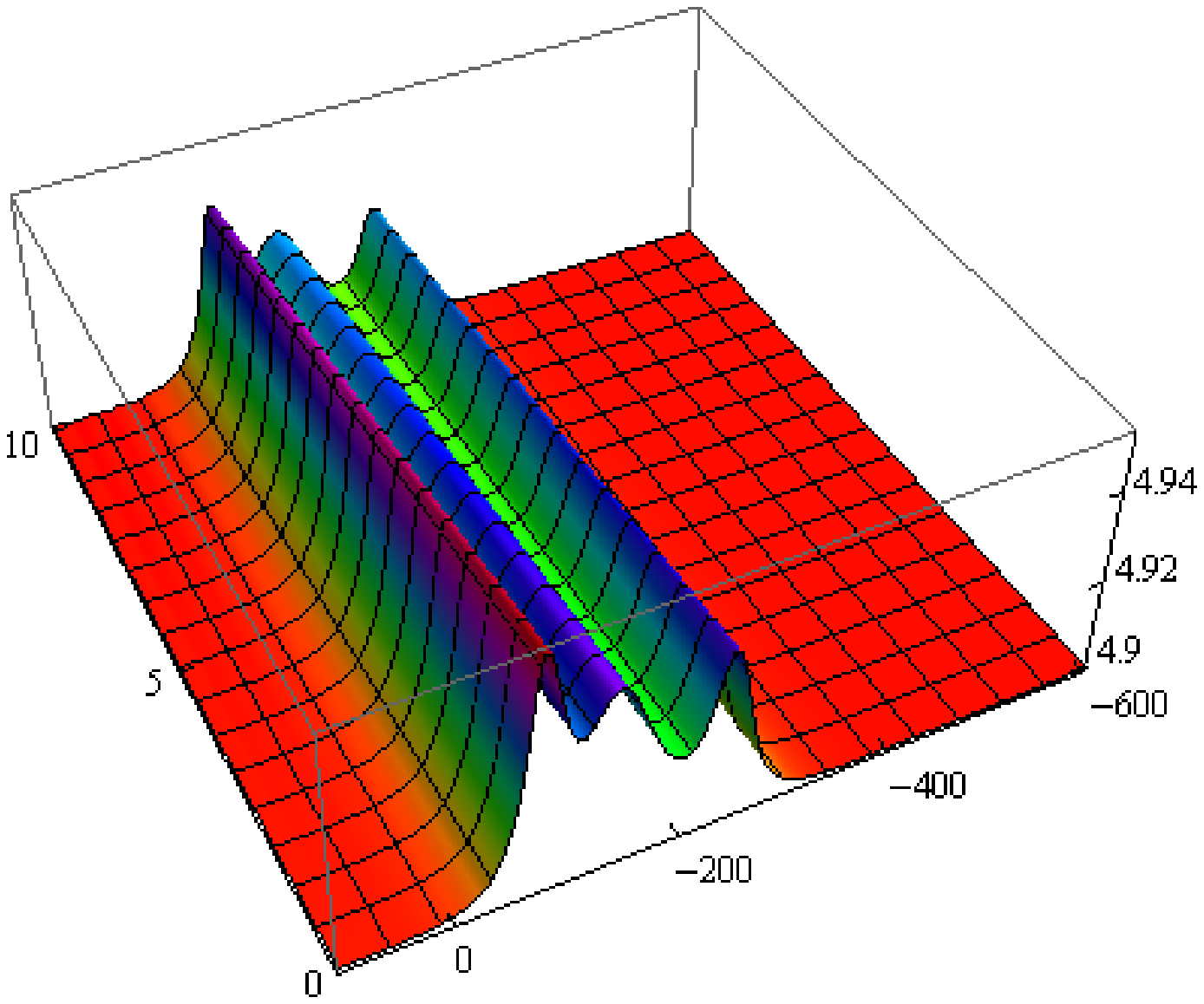} & \includegraphics[width=0.4\textwidth]{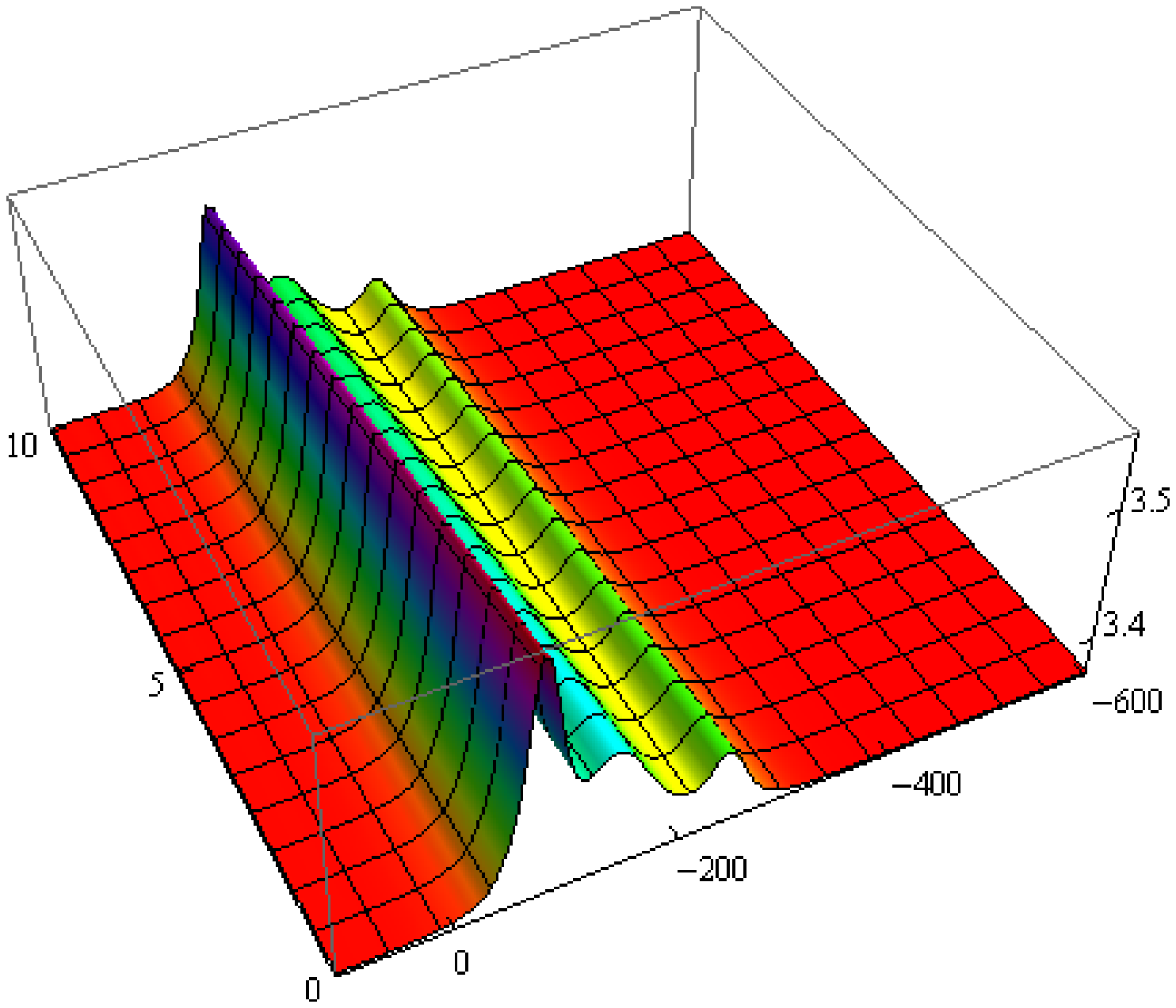}\\
\includegraphics[width=0.50\textwidth]{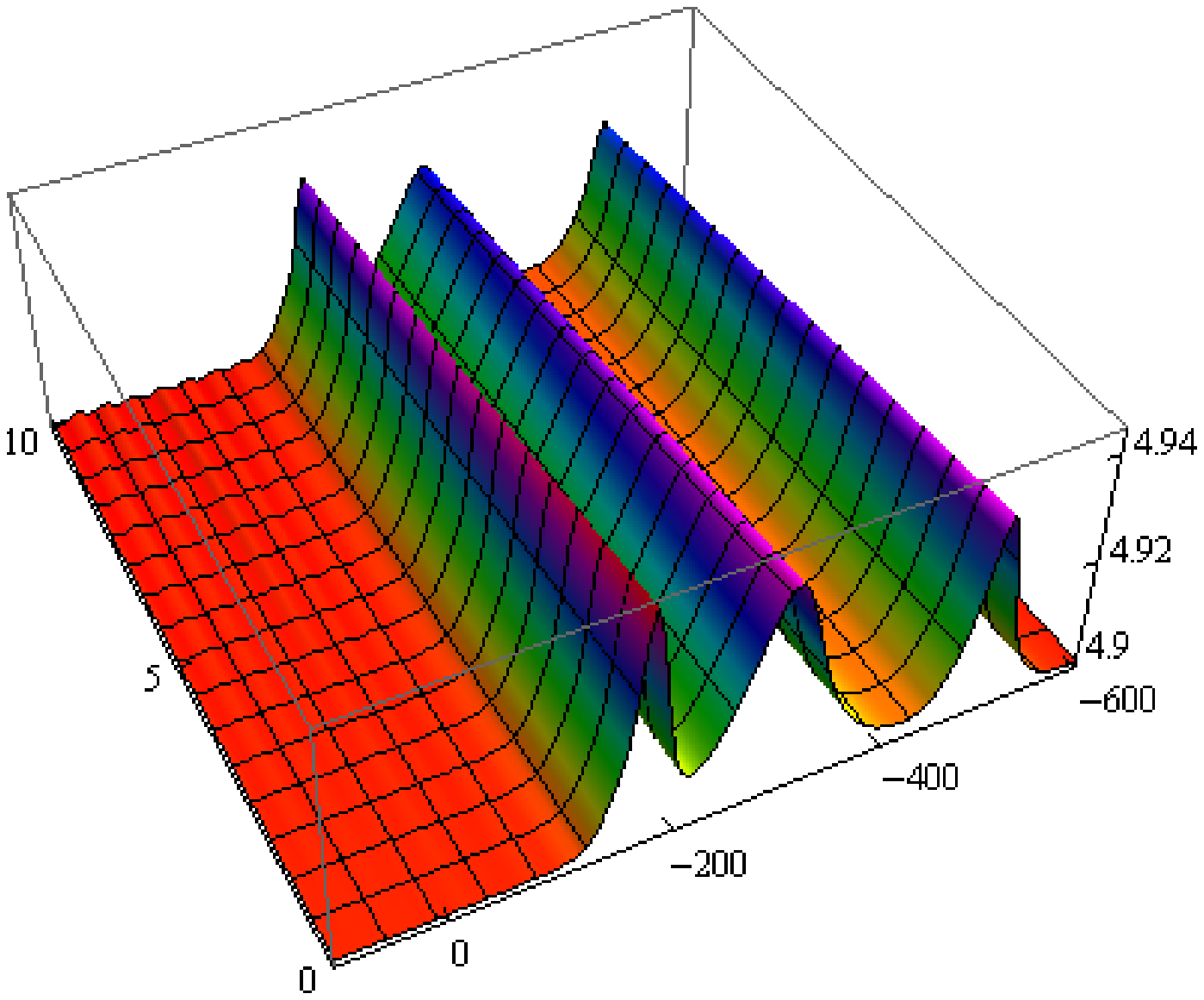} & \includegraphics[width=0.4\textwidth]{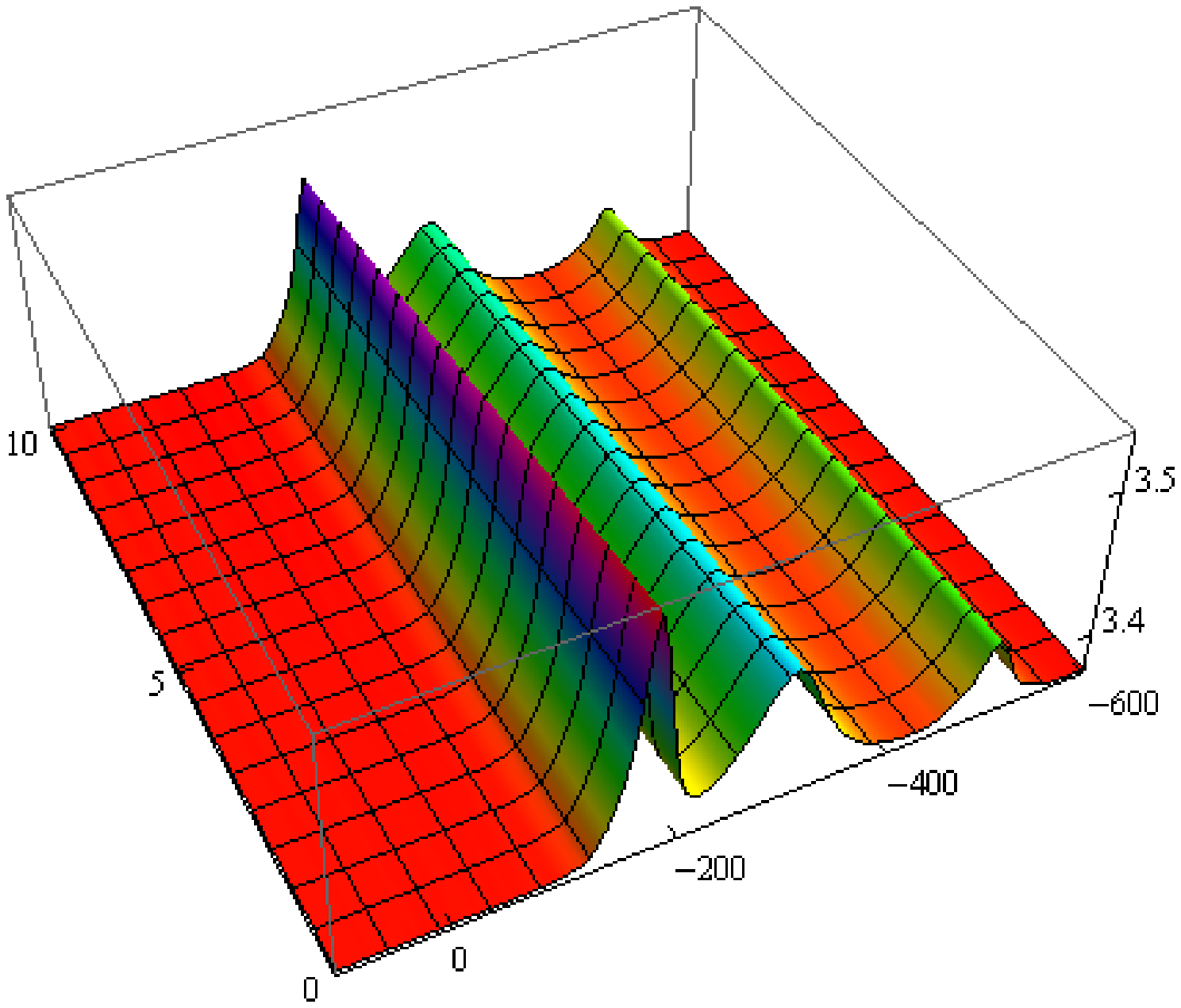}
\end{array} 
\put(-170,130){\fontsize{12}{12}$t$}
\put(-380,130){\fontsize{12}{12}$t$}
\put(-170,-30){\fontsize{12}{12}$t$}
\put(-380,-30){\fontsize{12}{12}$t$}
\put(-170,-190){\fontsize{12}{12}$t$}
\put(-380,-190){\fontsize{12}{12}$t$}
\put(-280,-60){\fontsize{12}{12}$y$}
\put(-60,-60){\fontsize{12}{12}$y$}
\put(-280,100){\fontsize{12}{12}$y$}
\put(-60,100){\fontsize{12}{12}$y$}
\put(-280,-210){\fontsize{12}{12}$y$}
\put(-60,-210){\fontsize{12}{12}$y$}
\put(-205,-91){\fontsize{12}{12}$b(60,y,t)$}
\put(-205,75){\fontsize{12}{12}$b(30,y,t)$} 
\put(-205,225){\fontsize{12}{12}$b(0,y,t)$}
\put(-410,225){\fontsize{12}{12}$a(0,y,t)$}
\put(-410,75){\fontsize{12}{12}$a(30,y,t)$}
\put(-410,-91){\fontsize{12}{12}$a(60,y,t)$}&
\end{array}$
\caption{Simulation, in the case $ N=3 $, of separation process depicted in figures \ref{Dinteract1}, \ref{Dinteract3} and characterized by (\ref{rho_k}). The 3D profiles for  $ a(x,y,t) $ and $ b(x,y,t) $ at different  $x$ positions are obtained through the numerical solution of system (\ref{hydro}) with $ x_{end}=60,\,t_{end}=1000,\,y_{end}=600 $ and initial/boundary data given by  (\ref{boundarydata1})--(\ref{bd}) along with (\ref{parameters1}).} \label{Figura5}
\end{figure}

\begin{figure}[tbp]
$
\begin{array}{c@{\hspace{1in}}c}
\begin{array}{cc}
\includegraphics[width=0.45\textwidth]{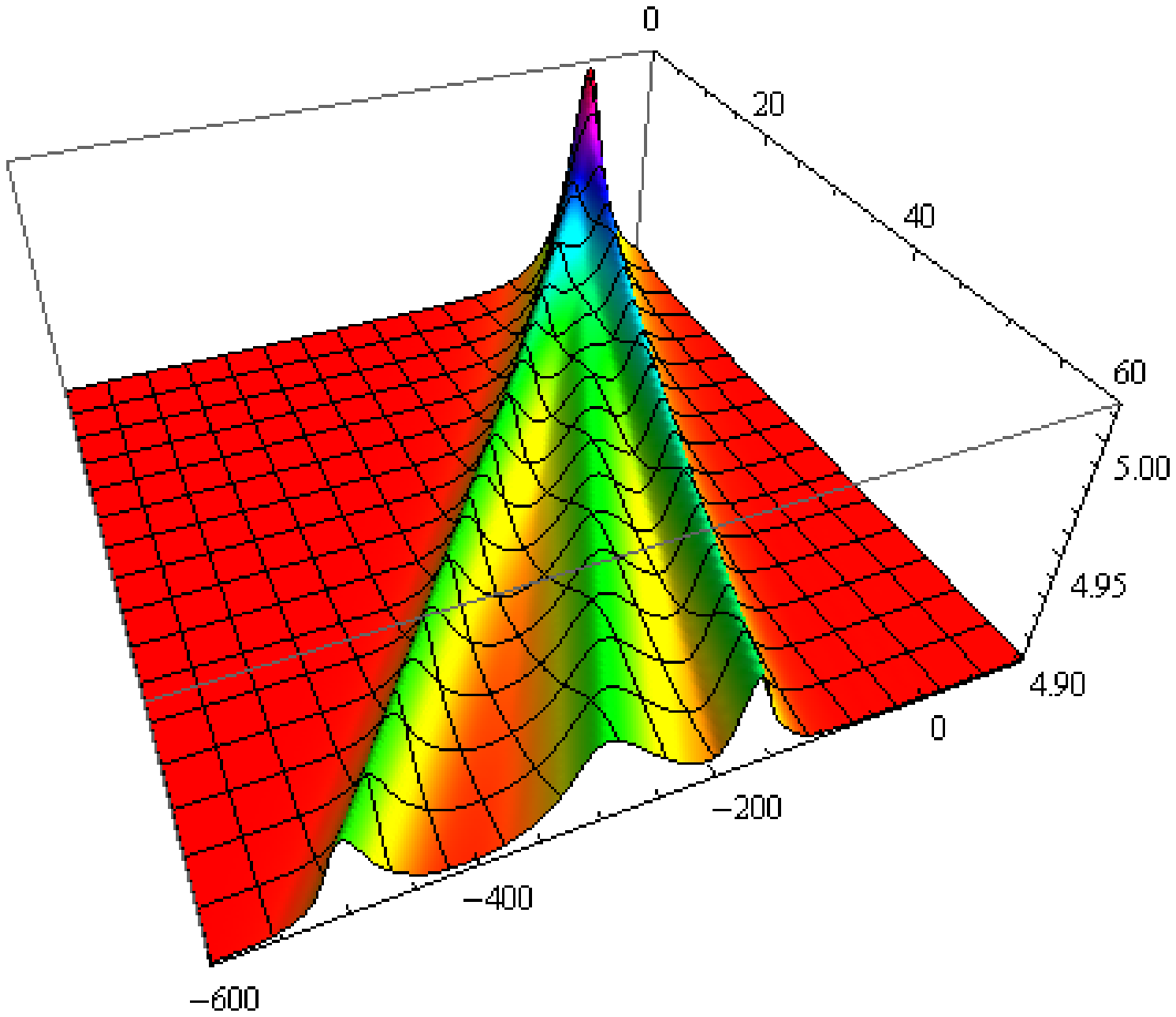} \,& \includegraphics[width=0.45\textwidth]{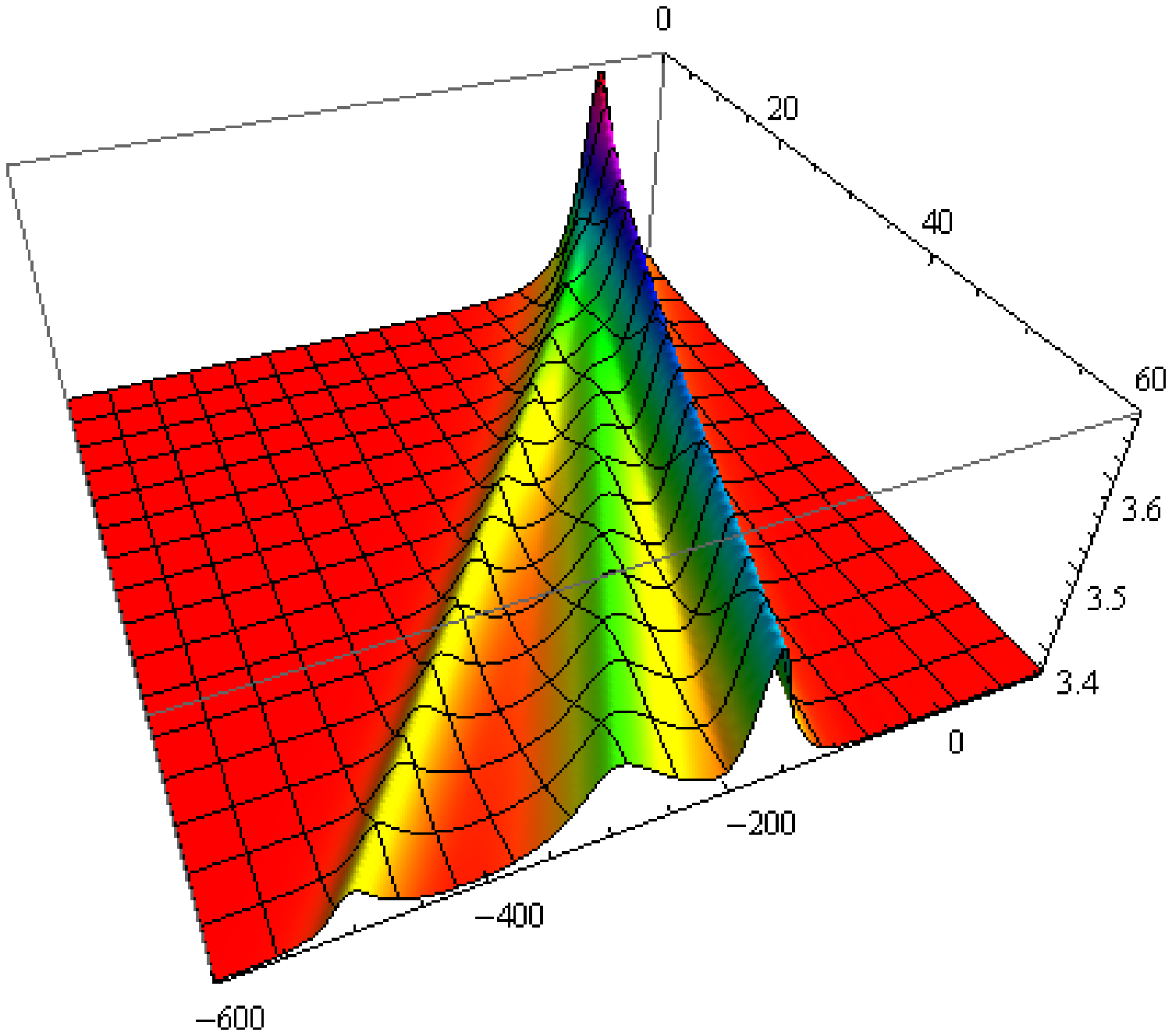} \medskip\\ 
\includegraphics[width=0.45\textwidth]{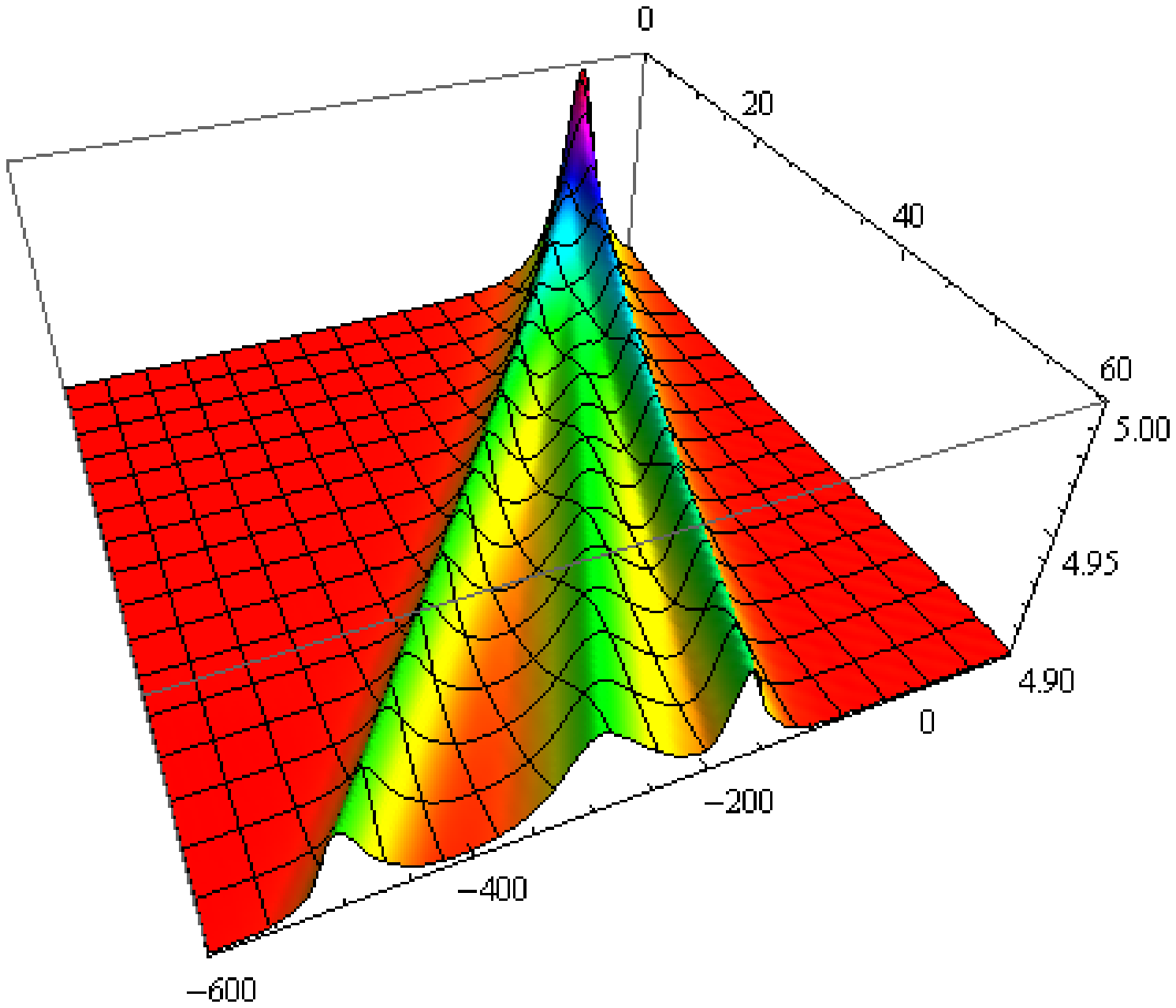} \,& \includegraphics[width=0.45\textwidth]{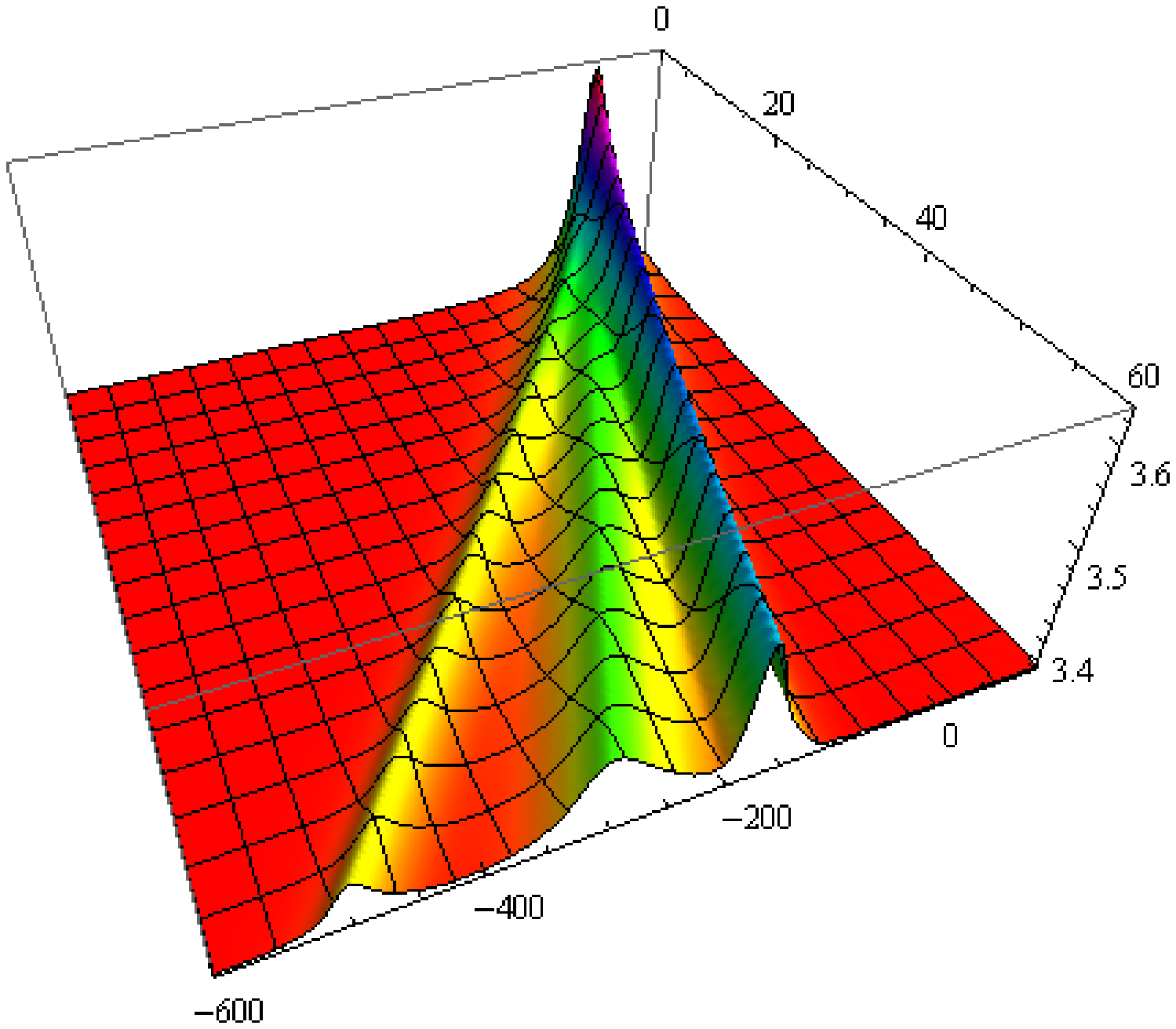}\\
\end{array} 
\put(-45,165){\fontsize{12}{12}$x$}
\put(-280,165){\fontsize{12}{12}$x$}
\put(-45,-30){\fontsize{12}{12}$x$}
\put(-280,-30){\fontsize{12}{12}$x$}
\put(-300,-160){\fontsize{12}{12}$y$}
\put(-90,-160){\fontsize{12}{12}$y$}
\put(-300,30){\fontsize{12}{12}$y$}
\put(-90,30){\fontsize{12}{12}$y$}
\put(-205,-15){\fontsize{12}{12}$b(x,y,5)$} 
\put(-205,175){\fontsize{12}{12}$b(x,y,0)$}
\put(-420,175){\fontsize{12}{12}$a(x,y,0)$}
\put(-420,-15){\fontsize{12}{12}$a(x,y,5)$}
\end{array}$
\caption{Simulation, in the case $ N=3 $, of separation process depicted in figures \ref{Dinteract1}, \ref{Dinteract3} and characterized by (\ref{rho_k}). The 3D profiles for  $ a(x,y,t) $ and $ b(x,y,t) $ at different times $t$ are obtained through the numerical solution of system (\ref{hydro}) with  the same initial/boundary data and parameters as in figure \ref{Figura5}.} \label{Figura6}
\end{figure}

\begin{figure}[tbp]
$
\begin{array}{c@{\hspace{1in}}c}
\centerline{\includegraphics[width=0.6 \textwidth]{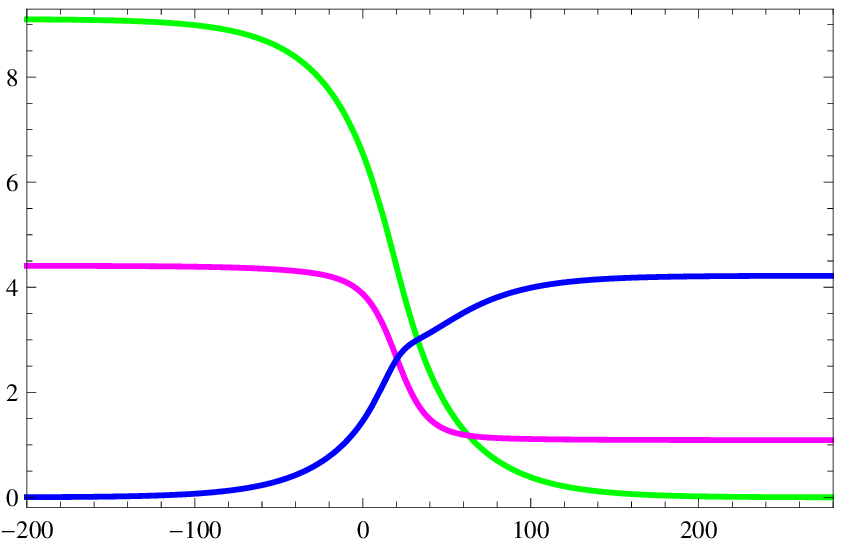}}
\put(-220,-10){\fontsize{12}{12}$y $}
\put(-300,100){\fontsize{12}{12}
$\Lambda_{2}(\alpha_{2})$} \put(-300,30){\fontsize{12}{12} $\Lambda_{3}(\alpha_{3})$} \put(-240,140){
\fontsize{12}{12}$\Lambda_{1}(\alpha_{1})$}
\end{array}
$
\caption{Interaction products defined in (\ref{interpro}) corresponding to the numerical solution depicted in figures \ref{Figura5} and \ref{Figura6}.}
 \label{Figura7}
\end{figure}

\begin{figure}[tbp]
$
\begin{array}{c@{\hspace{1in}}c}
\begin{array}{cc}
\includegraphics[width=0.42\textwidth]{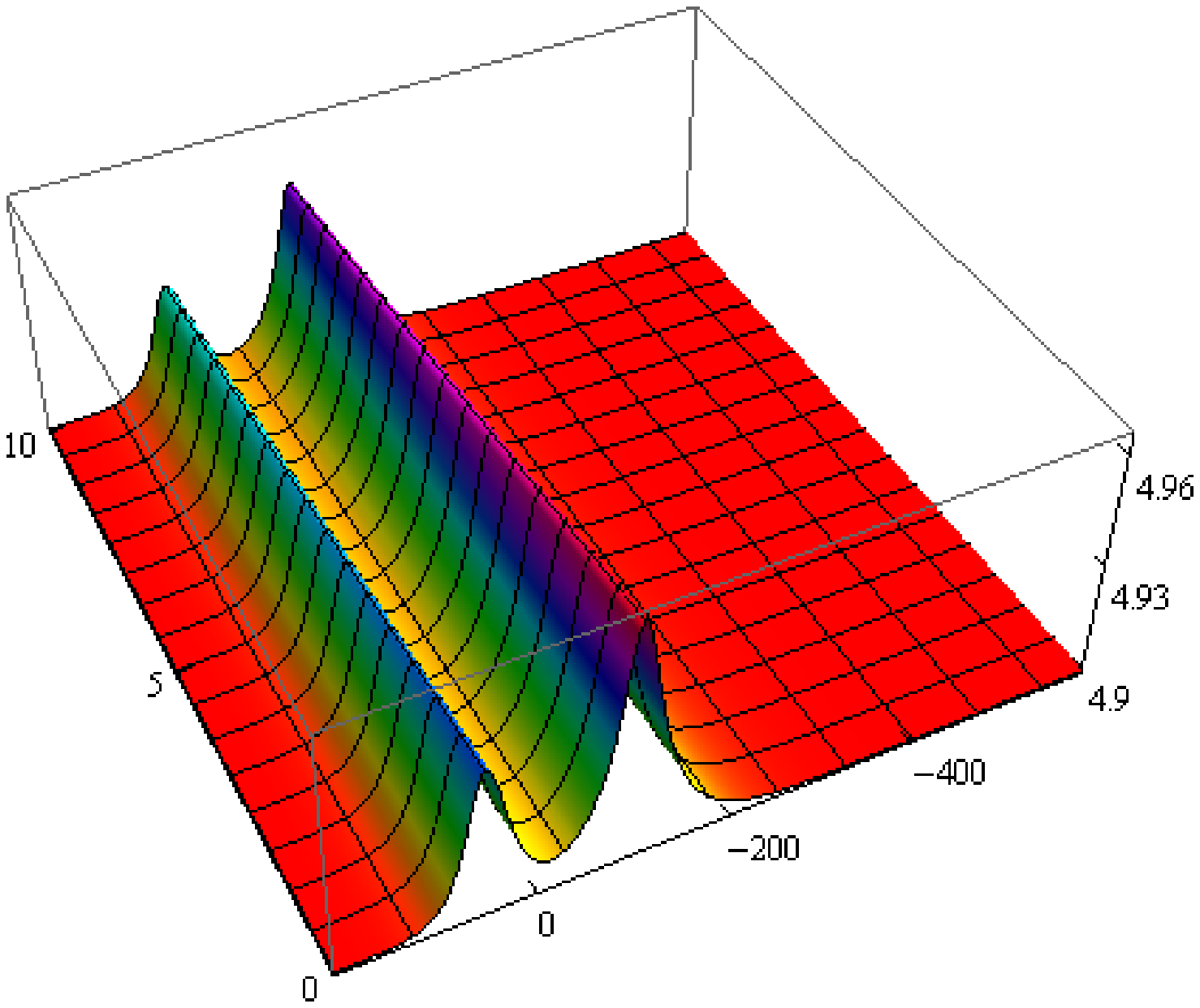} \,& \includegraphics[width=0.42\textwidth]{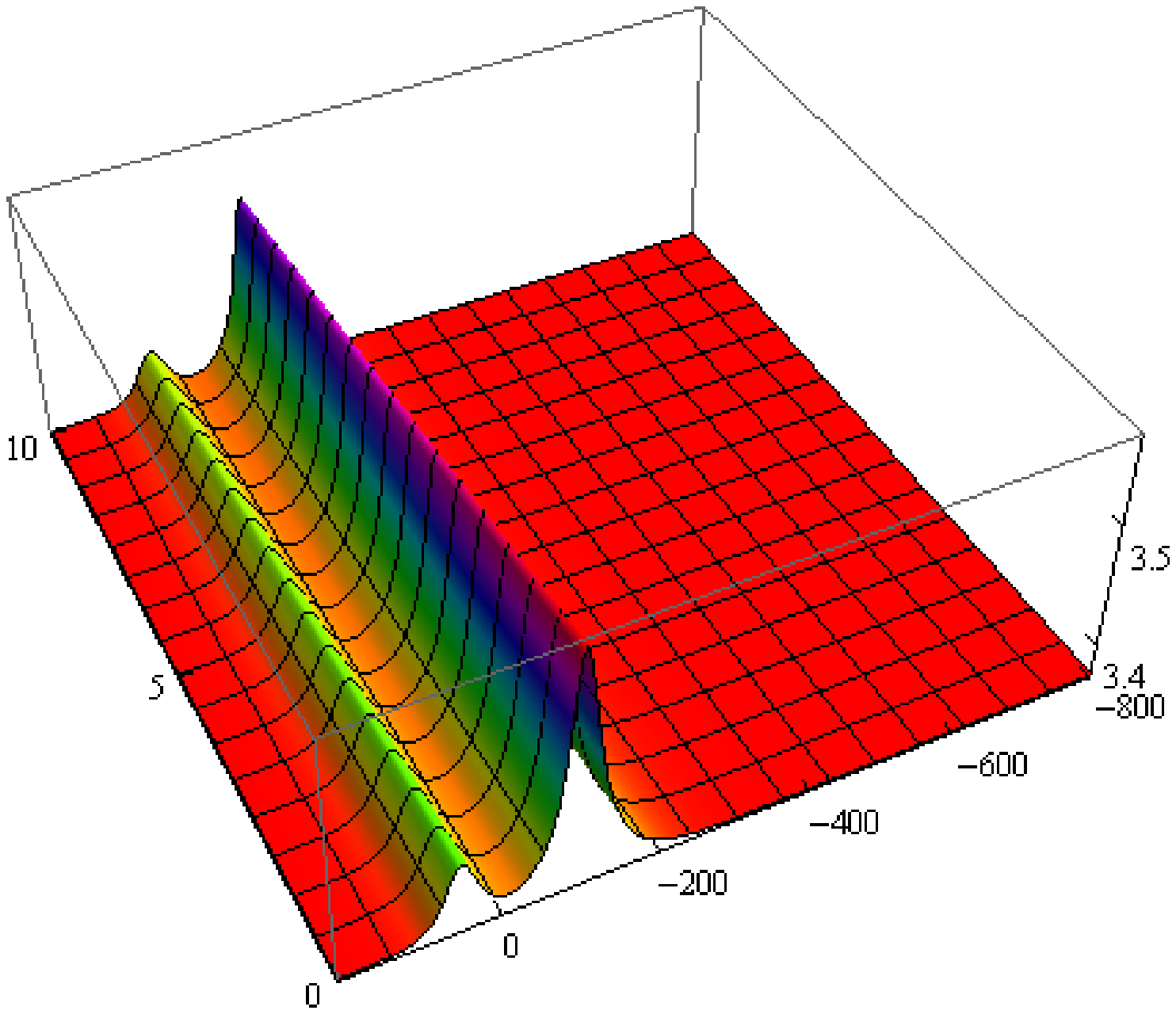} \medskip\\ 
\includegraphics[width=0.42\textwidth]{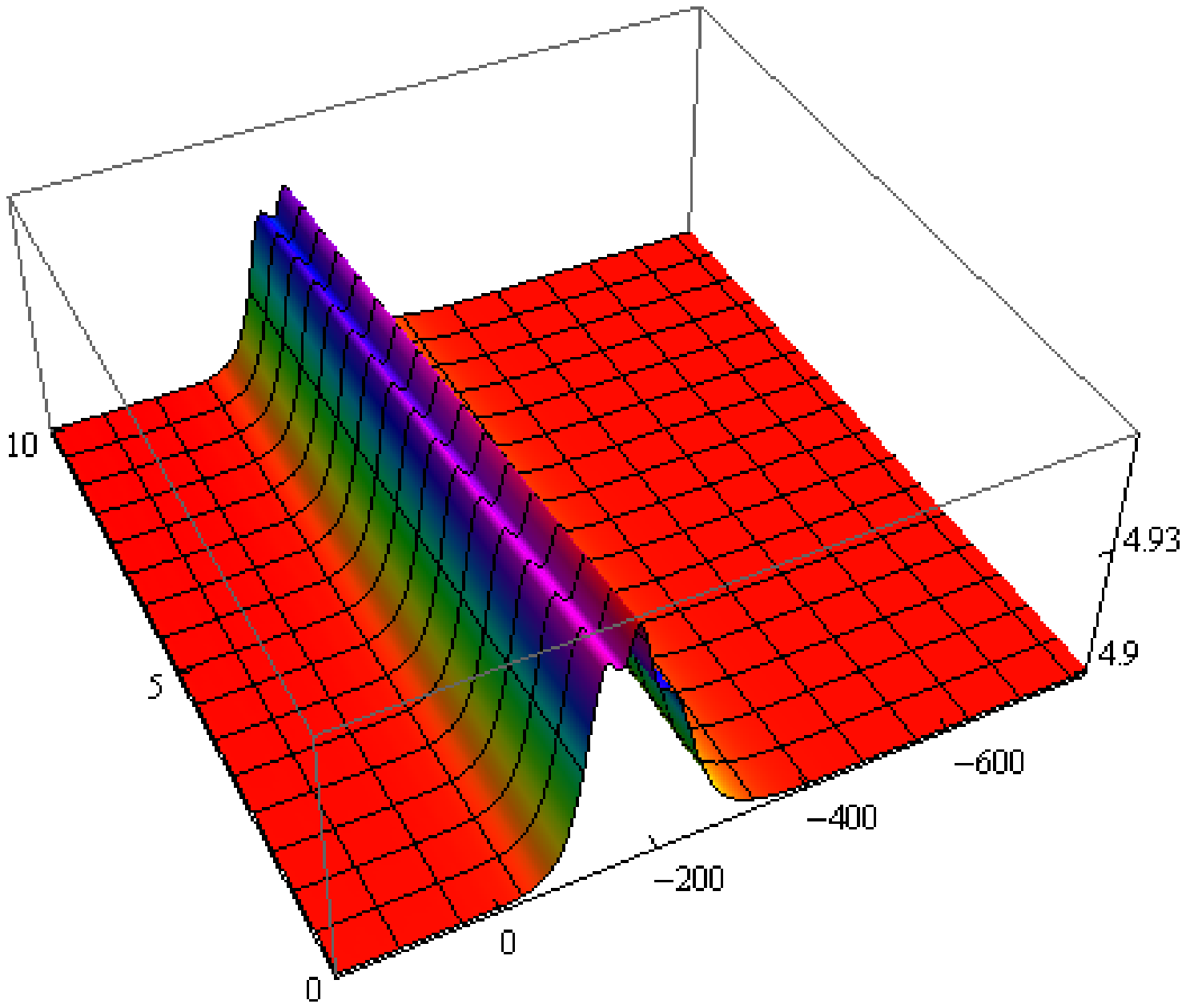}\, & \includegraphics[width=0.42\textwidth]{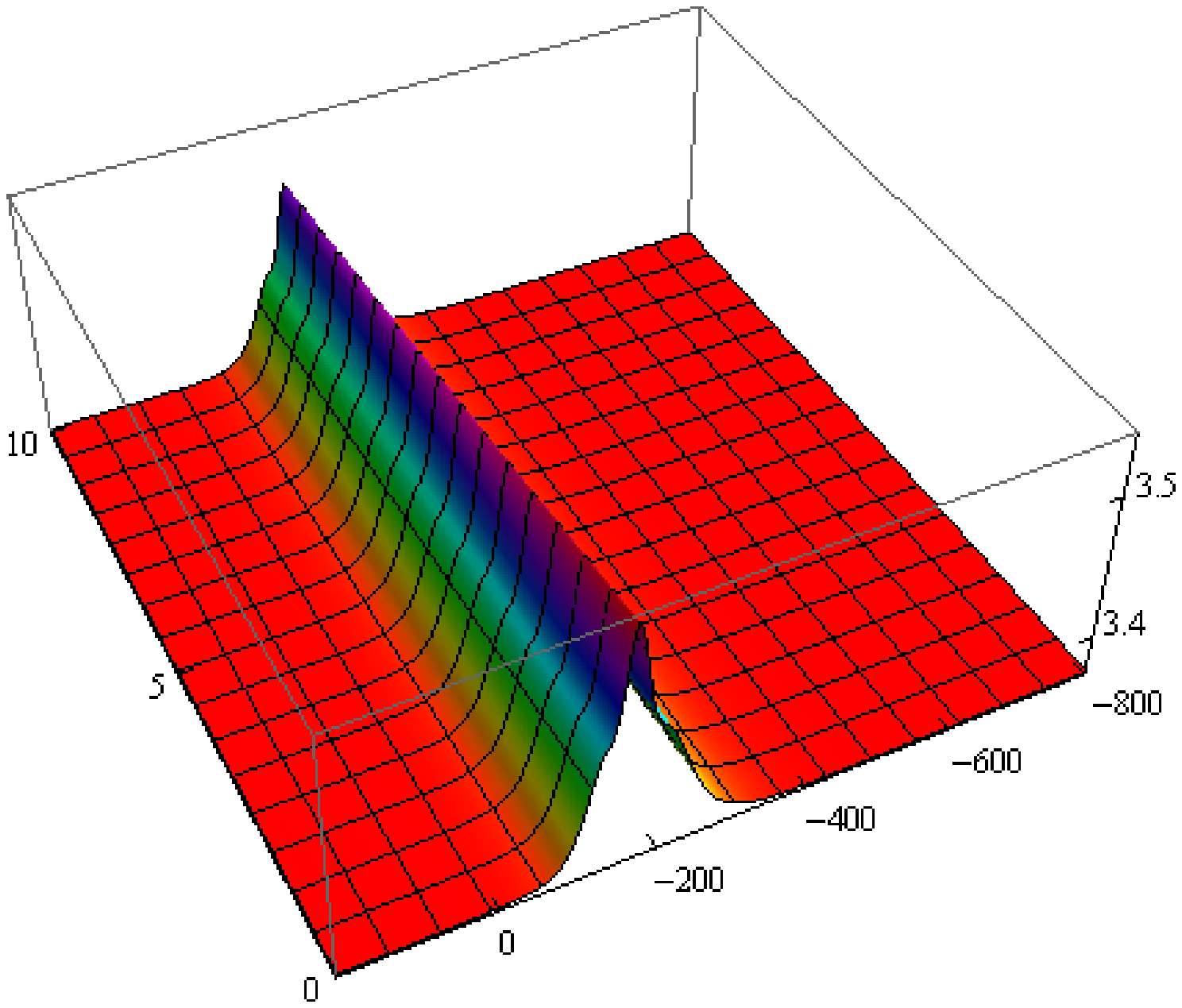}\medskip\\
\includegraphics[width=0.42\textwidth]{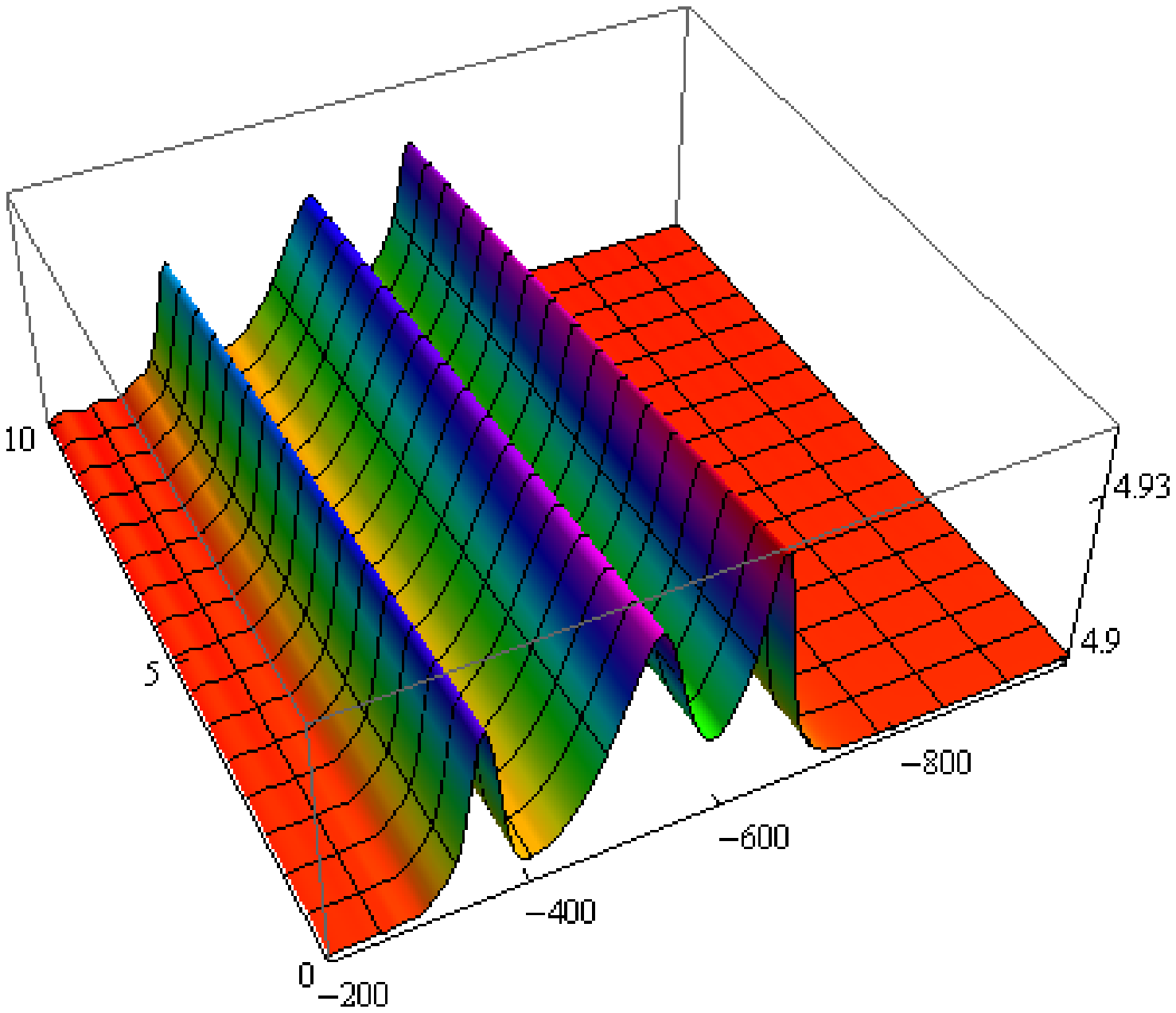}\, & \includegraphics[width=0.42\textwidth]{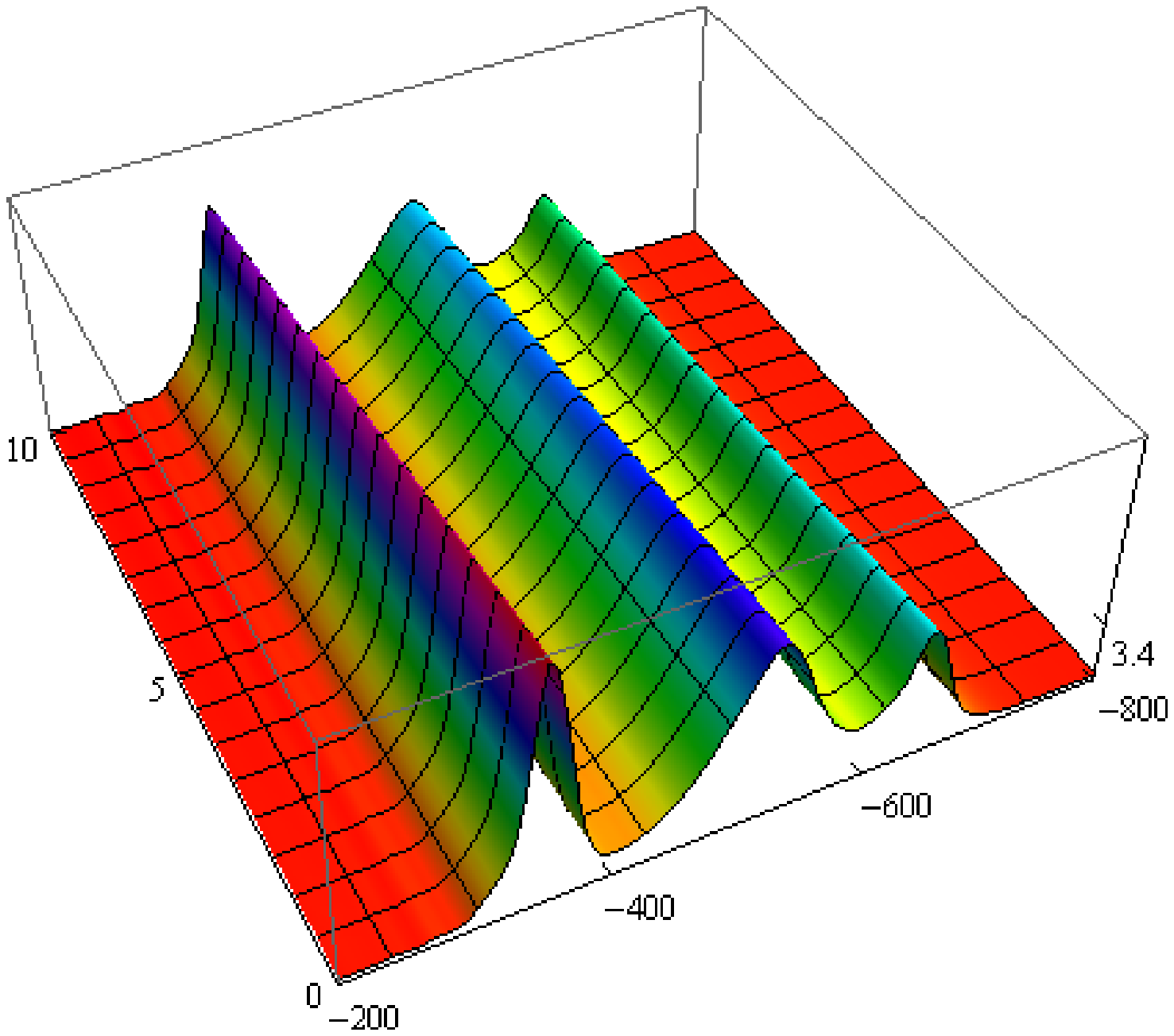}
\end{array} 
\put(-175,140){\fontsize{12}{12}$t$}
\put(-380,140){\fontsize{12}{12}$t$}
\put(-175,-30){\fontsize{12}{12}$t$}
\put(-380,-30){\fontsize{12}{12}$t$}
\put(-175,-200){\fontsize{12}{12}$t$}
\put(-380,-200){\fontsize{12}{12}$t$}
\put(-280,-60){\fontsize{12}{12}$y$}
\put(-60,-60){\fontsize{12}{12}$y$}
\put(-280,110){\fontsize{12}{12}$y$}
\put(-60,110){\fontsize{12}{12}$y$}
\put(-280,-235){\fontsize{12}{12}$y$}
\put(-60,-235){\fontsize{12}{12}$y$}
\put(-205,-100){\fontsize{12}{12}$b(80,y,t)$}
\put(-205,70){\fontsize{12}{12}$b(20,y,t)$} 
\put(-205,245){\fontsize{12}{12}$b(0,y,t)$}
\put(-400,245){\fontsize{12}{12}$a(0,y,t)$}
\put(-400,70){\fontsize{12}{12}$a(20,y,t)$}
\put(-400,-100){\fontsize{12}{12}$a(80,y,t)$}&
\end{array}$
\caption{Simulation, in the case $ N=3 $, of the interaction process depicted in figures \ref{Dinteract2}, \ref{Dinteract4} and characterized by (\ref{DATA_3}). The 3D profiles for  $ a(x,y,t) $ and $ b(x,y,t) $ at different $x$ positions are obtained through the numerical solution of system (\ref{hydro}) with $ x_{end}=90,\,t_{end}=1000,\,y_{end}=1000 $ and initial/boundary data given by (\ref{boundarydata1})--(\ref{bd}) along with (\ref{parameters2}).} \label{Figura8}
\end{figure}

\begin{figure}[tbp]
$
\begin{array}{c@{\hspace{1in}}c}
\begin{array}{cc}
\includegraphics[width=0.45\textwidth]{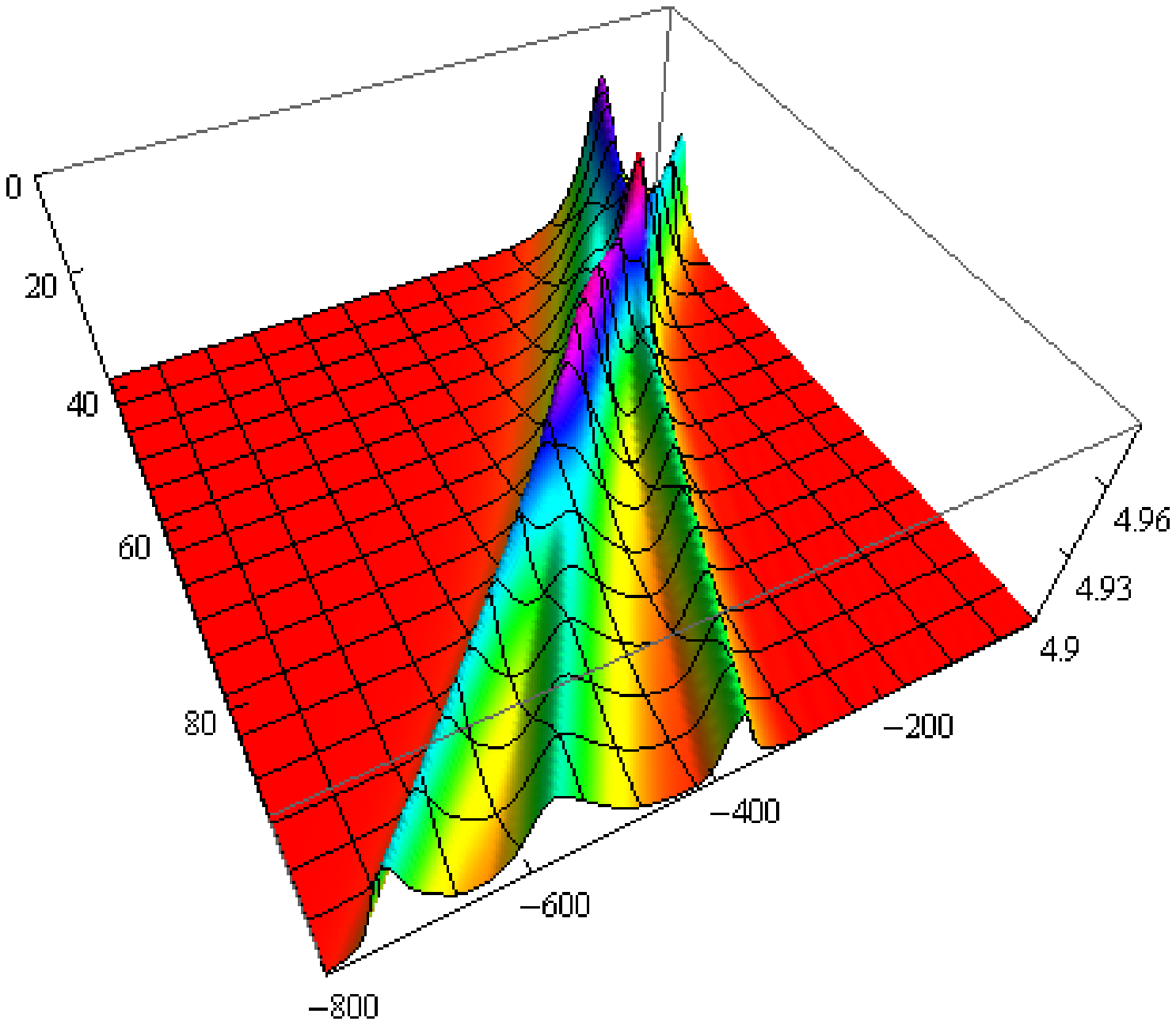} \,& \includegraphics[width=0.45\textwidth]{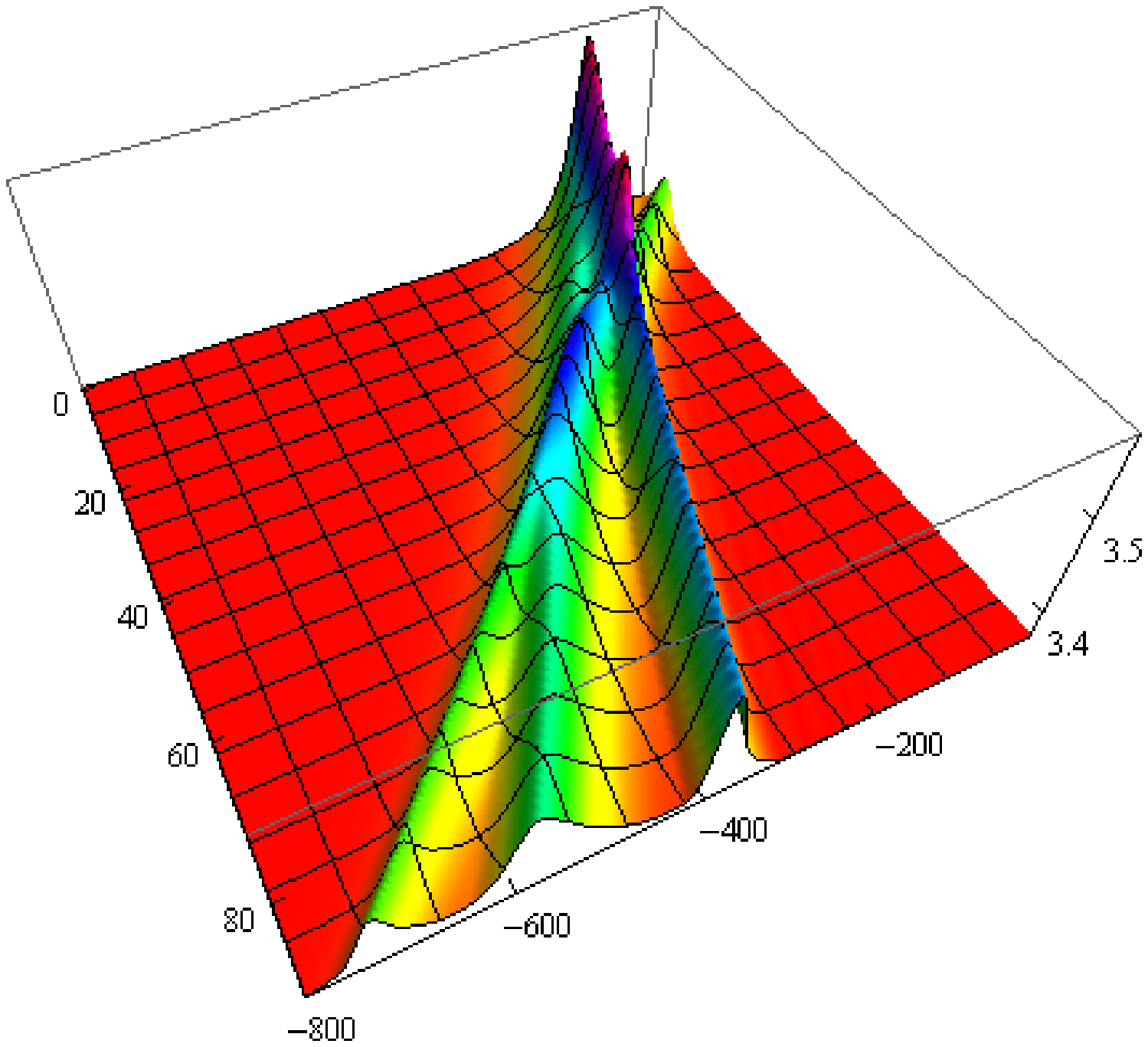} \medskip\\ 
\includegraphics[width=0.45\textwidth]{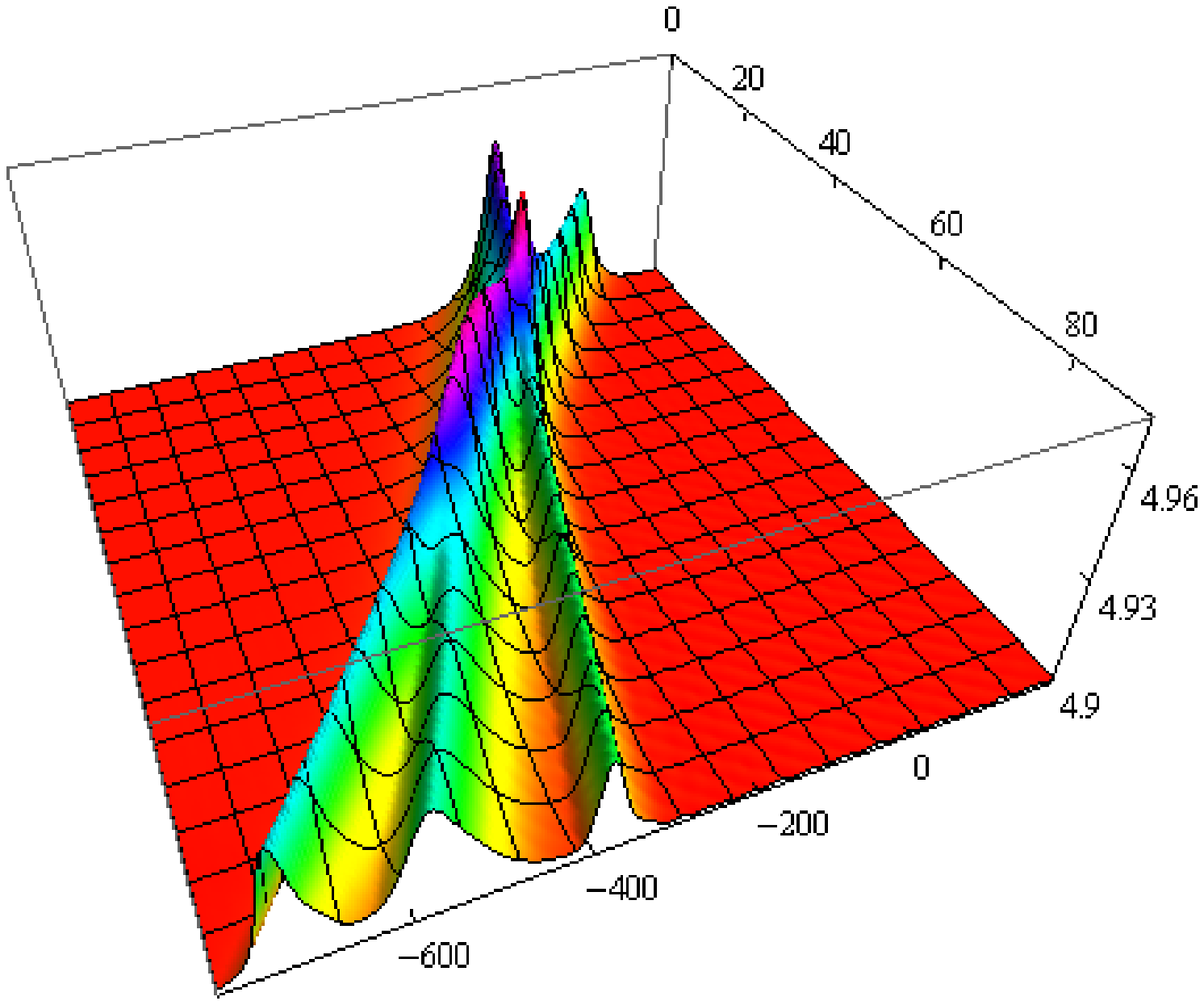} \,& \includegraphics[width=0.45\textwidth]{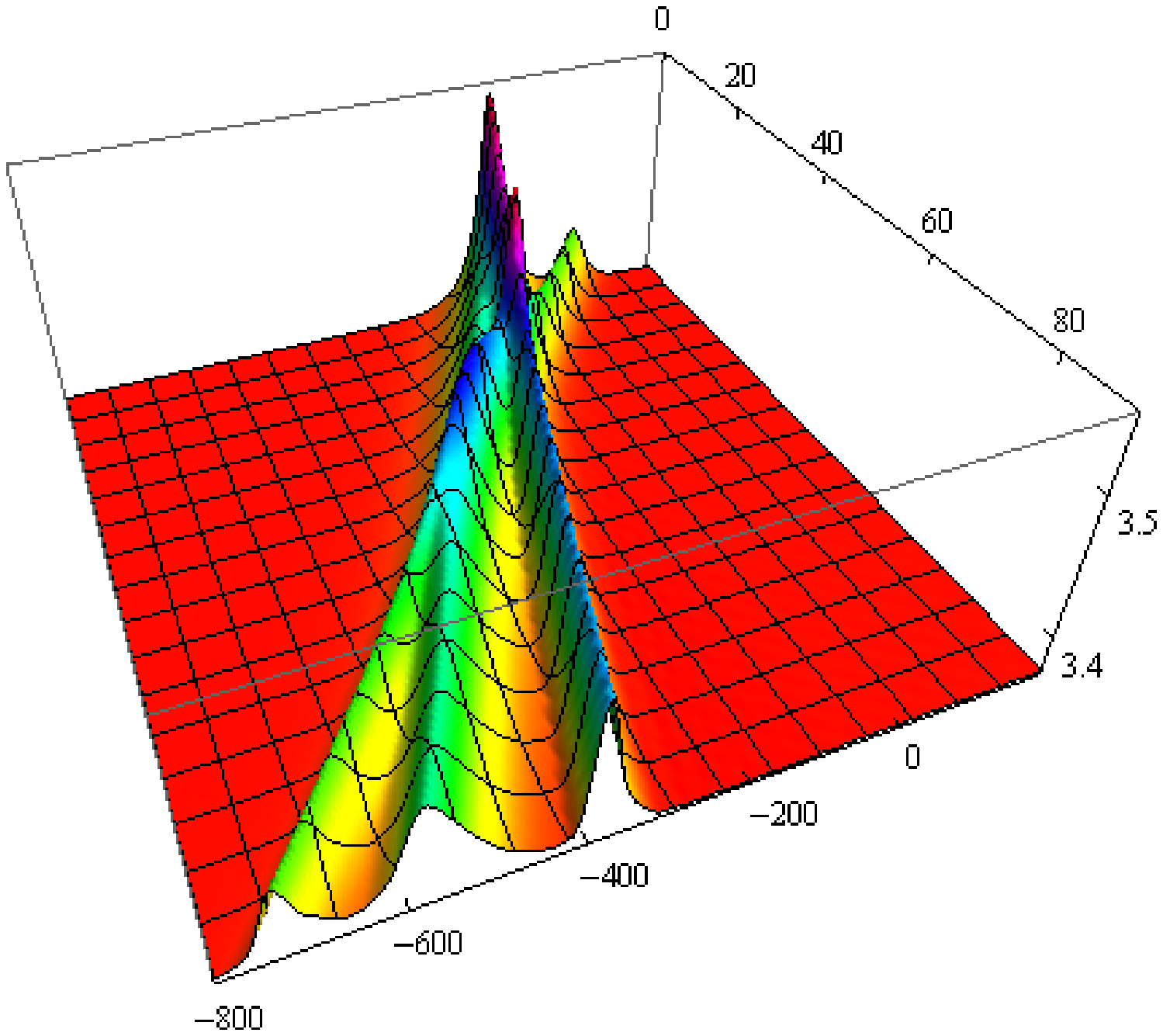}\\
\end{array} 
\put(-45,165){\fontsize{12}{12}$x$}
\put(-280,165){\fontsize{12}{12}$x$}
\put(-45,-30){\fontsize{12}{12}$x$}
\put(-270,-42){\fontsize{12}{12}$x$}
\put(-300,-175){\fontsize{12}{12}$y$}
\put(-90,-170){\fontsize{12}{12}$y$}
\put(-300,30){\fontsize{12}{12}$y$}
\put(-90,30){\fontsize{12}{12}$y$}
\put(-205,-19){\fontsize{12}{12}$b(x,y,5)$} 
\put(-205,180){\fontsize{12}{12}$b(x,y,0)$}
\put(-420,175){\fontsize{12}{12}$a(x,y,0)$}
\put(-420,-25){\fontsize{12}{12}$a(x,y,5)$}
\end{array}$
\caption{Simulation, in the case $ N=3 $, of the interaction process depicted in figures \ref{Dinteract2}, \ref{Dinteract4} and characterized by (\ref{DATA_3}). The 3D profiles for  $ a(x,y,t) $ and $ b(x,y,t) $ at different times $ t$ are obtained through the numerical solution of system (\ref{hydro}) with the same initial/boundary data and parameters as in figure \ref{Figura8}.}  \label{Figura9}
\end{figure}
\begin{figure}[tbp]
$
\begin{array}{c@{\hspace{1in}}c}
\begin{array}{cc}
\includegraphics[width=0.45\textwidth]{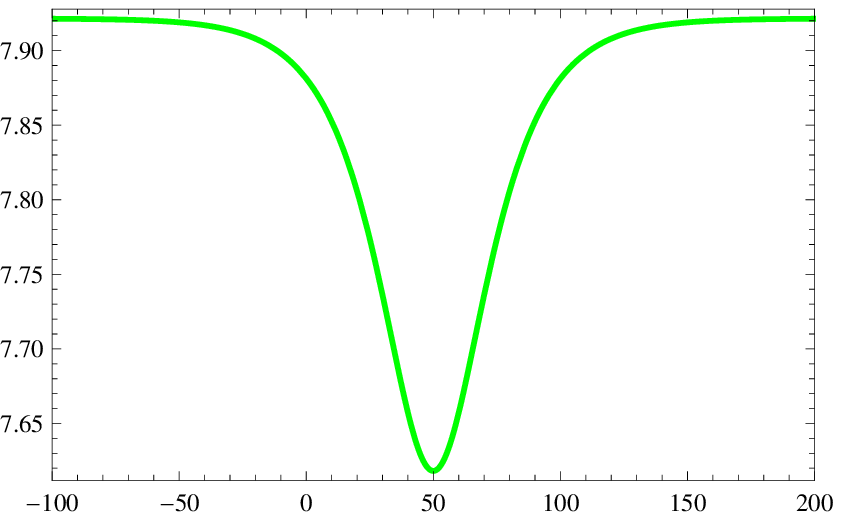} \,& \includegraphics[width=0.45\textwidth]{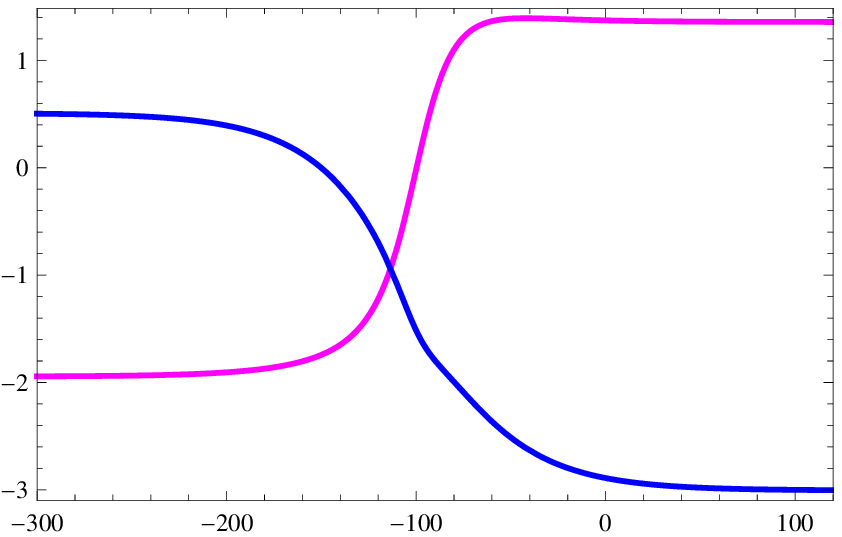} \\
\end{array} 
\put(-320,-70){\fontsize{12}{12}$y $}
\put(-100,-70){\fontsize{12}{12}$y $}
\put(-180,-10){\fontsize{12}{12}
$\Theta_{2}(\alpha_{2})$} \put(-180,50){\fontsize{12}{12} $\Theta_{3}(\alpha_{3})$} \put(-390,36){
\fontsize{12}{12}$\Theta_{1}(\alpha_{1})$}
\end{array}
$
\caption{Interaction products defined in (\ref{interpro_III}), (\ref{interpro_IV}) corresponding to the numerical solution depicted in figures \ref{Figura8} and \ref{Figura9}.}
 \label{Figura10}
\end{figure}

\section{Numerical results: case $ N=3 $}\label{numerical}
In this section, in order to validate the analytical results previously obtained as well as to get a deeper insight into the interaction/separation processes described hitherto, we integrate numerically the set of equations (\ref{hydro}). To this aim  the $ x $   coordinate is considered as the "evolution" variable so that the computation is performed for $ x\in \left[0,x_{end}\right]$, over the domain $ (y,t)\in \left[-y_{end},y_{end}\right] \times \left[0,t_{end}\right] $.

According to (\ref{a-b}) the "initial" data for $ a $ and $ b $ are obtained from
\begin{equation} \label{boundarydata1}
a(0,y,t)=\sum_{m=1}^{3} R^{m}(0,y,t),\quad\quad b(0,y,t)=\prod_{m=1}^{3}R^{m}(0,y,t)
\end{equation}
where
\begin{equation} \label{boundarydata}
R^{m}(0,y,t)=c_{m}+d_{m}\exp\left(-\frac{t}{\epsilon_{m}}\right) {\rm sech}\left(h_{m}(y-y^0_{m})\right),\quad ( m=1,2,3 )
\end{equation}
and $y^0_{m}\in \left[-y_{end},y_{end}\right] $, $ c_{m},\,d_{m},\,\epsilon_{m},\,h_{m} $  are constants.
 
As far as the "boundary" data  $ a(x,y,0) $ and $ b(x,y,0) $  are concerned, since the relations (\ref{sol}) evaluated at $ t=0 $ represent the general solution of the  homogeneous system of hydrodynamic type (\ref{chrom2}) in $ 1+1 $ variables \cite{Tsarev1,Tsarev2}, then in the numerical procedure at hand we assume the following "boundary" conditions
\begin{equation}
 a(x,y,0)=\sum_{m=1}^{3} \omega^{m}(x,y),\quad\quad b(x,y,0)=\prod_{m=1}^{3}\omega^{m}(x,y) \label{bd}
\end{equation}
where $\omega^{m}(x,y)$ are given by numerical integration of (\ref{chrom2}) for  $ x\in \left[0,x_{end}\right]$ over the spatial domain $ y\in \left[-y_{end},y_{end}\right] $ with initial conditions $\omega^{m}(0,y)= R^{m}(0,y,0)$.

In figures \ref{Figura5} and \ref{Figura6}, by considering the set of parameteres
\begin{equation}\label{parameters1}
\fl \eqalign{
 c_{1}=2.5,\, c_{2}=1.5,\,c_{3}=0.9 , \quad d_{1}=0.035,\,d_{2}=0.038, \, d_{3}=0.040, \cr
 h_{1}=0.055, \, h_{2}=0.030,\,h_{3}=0.080 ,\quad \epsilon_{1}=\epsilon_{2}=\epsilon_{3}=0.010 ,\quad y^0_{1}=y^0_{2}=y^0_{3}=20, }
\end{equation}
we show spatial and temporal evolution of an initial pulse characterized by (\ref{boundarydata1}), (\ref{boundarydata}) and (\ref{bd}).
It should be noticed that the behaviours of the field variables $ a(x,y,t) $ and $ b(x,y,t) $  support the qualitative analysis depicted in figures \ref{Dinteract1} and \ref{Dinteract3}. In fact the choice $ y^0_{1}=y^0_{2}=y^0_{3} $ simulates a pulse localized at $ x=0 $ in a stripe of the $(y,t)- $plane (as illustrated in figures \ref{Figura5} and \ref{Figura6}) which after a finite distance $ x $ (interaction region) separates into three simple waves travelling along different characteristic curves $ C^{(k)} $ and become separate by regions of constant states.
Moreover in figure \ref{Figura7} we plot the control parameters $ \Lambda_{(k)} $ associated to (\ref{boundarydata}) and (\ref{parameters1}) for the simple waves deformation in the interaction region. In this case the boundary data  provide $ y_{1}\simeq -100 $ and  $ y_{2}\simeq 140 $ therefore, according to (\ref{interpro}), the profiles of the characteristic curves $ C^{(1)} $ and $ C^{(3)} $  originating at $ y_2 $  and $ y_1 $ respectively, do not distort being $ \Lambda_{1}(y_{2}) =\Lambda_{3}(y_{1})=0$. Moreover, as predicted by (\ref{interpro}), all the $ \Lambda_{(k)} $ become constant for $ \alpha_{k}\notin \left[y_1, y_2\right] $.

Next we choose the following set of parameters
\begin{equation}\label{parameters2}
\fl \eqalign{
  c_{1}=2.5,\, c_{2}=1.5,\,c_{3}=0.9 , \quad d_{1}=0.040,\,d_{2}=0.035, \, d_{3}=0.030, \cr
 h_{1}=0.055, \, h_{2}=0.030,\,h_{3}=0.080 ,\; \epsilon_{1}=\epsilon_{2}=\epsilon_{3}=0.010 ,\; y^0_{1}=50, \, y^0_{2}=y^0_{3}=-100.}
\end{equation}
in order to simulate a simple wave and a pulse localized at $ x=0 $ in different stripes of the $(y,t)- $plane (as illustrated in figures \ref{Figura8} and \ref{Figura9}).
In this case, by integrating the system of PDEs (\ref{hydro}), we notice that the behaviours of the field variables $ a(x,y,t) $ and $ b(x,y,t) $ confirm the qualitative analysis depicted in figures \ref{Dinteract2} and \ref{Dinteract4}.  
In particular figures \ref{Figura8} and \ref{Figura9} show that
the simple wave, travelling along the fastest characteristic curve $ C^{(1)} $,  interacts with the pulses travelling along $ C^{(2)} $ and $ C^{3)} $, affects their profiles and emerges again as simple wave. Moreover, after a finite distance $ x $, the initial pulse  separates into three simple waves.
Finally in figure \ref{Figura10} we plot the control parameters $ \Theta_{(k)} $ associated to (\ref{boundarydata}) and (\ref{parameters2}) for the simple waves deformation. 
In this case the boundary data  provide $ y_{1}\simeq 0 $,  $ y_{2}\simeq 50 $, $ y_{3}\simeq -160 $  and $ y_{4}\simeq -40 $ therefore, according to (\ref{interpro_III}), (\ref{interpro_IV}),  $ \Theta_{(2)} $ and $ \Theta_{(3)} $ become constant for increasing values of $ \alpha_2 $ and  $ \alpha_3 $  whereas $ \Theta_{(1)} $ is constant if  $ \alpha_{1}\notin \left[y_1, y_2\right] $.

\section{Conclusion}\label{conclusion}
The existence of Riemann invariants plays a fundamental role in studying two dimensional nonlinear wave interactions as well as for some physically interesting examples of multi--component  hydrodynamic systems as in chromatography or electrophoresis models which can be also written in a diagonal form. However in multi dimensional case nonlinear wave interactions are not so deeply understood because in such a case the governing systems do not possess, in general, Riemann invariants. Nevertheless, in the theory of integrable multi dimensional quasilinear systems of first order, every hydrodynamic reduction admittes Riemann invariants. Since any integrable multi dimensional quasilinear system of first order possesses infinitely many multi component hydrodynamic reductions, one can select the appropriate hydrodynamic reduction according to given boundary conditions or Cauchy initial problems. By this reason in this paper we considered a three dimensional two component integrable quasilinear system of first order which has infinitely many hydrodynamic reductions. For simplicity we restricted our attention to multi-component reductions which already are known as the chromatography system, because in this particular case we already know how to construct a general solution. For such a model we gave an exact analytical description of three dimensional nonlinear wave interactions.

Such a theoretical problem was usually considered for two-dimensional hyperbolic models \cite{cur1,cur2,SeymourV,CurroFusco}, while, in the case of more space variables, for special evolution processes ruled by an
auxiliary $2\times 2$ hyperbolic subsystem \cite{CurroFusco02,CurroFusco13}.

The crucial point of the procedure at hand is the existence of $N$
Riemann invariants for each particular hydrodynamic reduction (see details in \cite{maksjmp}) of three dimensional quasilinear system (\ref{hydro}). Therefore the method can be applied to the class of  semi-Hamiltonian
homogeneous systems of hydrodynamic type which admit a diagonalized form \cite{Tsarev1}.
In this case  a general solution can be  obtained by means of the generalized hodograph method \cite{Tsarev2}.

Although the study developed herein can be performed for any diagonalizable semi-Hamiltonian
homogeneous model which can be solved by the generalized hodograph method, the procedure was illustrated for two commuting systems (\ref{chrom1}) and (\ref{chrom2}) associated to the three dimensional system (\ref{hydro}). Two different situations were illustrated. First we described the evolution of $N$ waves initially localized in a closed interval, next we studied the interaction between $N-1$ pulses  and a further single wave. In both cases the resulting waves behave like simple waves and after the interaction region they are distorted. The amount of the produced distortion is  analytically computed  through the $\Lambda_k (\alpha_k)$ and $\Theta_k (\alpha_k)$ terms which depend on the initial data given, respectively, by (\ref{rho_k}) and (\ref{DATA_3}). As it was pointed out in section \ref{inivalue}, although in section \ref{Nwave} we studied nonlinear wave interactions concerning the examples characterized by the initial/boundary conditions (\ref{rho_k}) and (\ref{DATA_3}), we remark that our procedure can be in principle applied to describe hyperbolic nonlinear wave interactions corresponding to any set of initial/boundary data.  

Finally, in order to validate the analytical results herein obtained, the numerical integration of the governing system (\ref{hydro}) was performed. The resulting three dimensional figures confirm the behaviours which have been analytically described by means of the proposed procedure.

\ack{The authors sincerely thank the anonymous Referees for the constructive comments raised during the review stage.
MVP's work was partially supported by the Russian Science Foundation (grant No. 15-11-20013). CC and NM thank the Italian National Group of Mathematical Physics (GNFM-INDAM) for supporting the present research.}

\end{document}